\documentclass[11pt,preprint]{aastex}
\usepackage{amsmath}
\newcommand{\vecs}{\mbox{\boldmath $s$}}

\newcommand{\veck}{\mbox{\boldmath $k$}}
\newcommand{\vecn}{\mbox{\boldmath $n$}}
\newcommand{\veci}{\mbox{\boldmath $i$}}
\newcommand{\vecj}{\mbox{\boldmath $j$}}

\newcommand{\au}{\,{\rm AU}}

\newcommand{\Myr}{\,{\rm Myr}}
\newcommand{\ME}{\ensuremath{\mathrm{M_{\Earth}}}}
\newcommand{\epsJ}{\ensuremath{\varepsilon_{\mathrm{J}}}}
\newcommand{\epsS}{\ensuremath{\varepsilon_{\mathrm{S}}}}

\begin{document}
\title{Tilting Saturn without tilting Jupiter: Constraints on giant planet migration}
\author{R.~Brasser\altaffilmark{1} and Man~Hoi~Lee\altaffilmark{2}}
\altaffiltext{1}{Earth-Life Science Institute, Tokyo Institute of Technology, Meguro, Tokyo 152-8551, Japan}
\altaffiltext{2}{Department of Earth Sciences and Department of Physics, The University of Hong Kong, Pokfulam Road, Hong Kong}

\begin{abstract}
The migration and encounter histories of the giant planets in our Solar System can be constrained by the obliquities of Jupiter
and Saturn. We have performed secular {simulations with imposed migration} and $N$-body simulations with {planetesimals} to
study the expected obliquity distribution of migrating planets with initial conditions resembling those of the smooth migration model,
the resonant Nice model and two models with five giant planets initially in resonance (one compact and one loose configuration). For 
smooth migration, the secular spin-orbit resonance mechanism can tilt Saturn's spin axis to the current obliquity if the product of 
the migration time scale and the orbital inclinations is sufficiently large (exceeding 30~Myr~deg). For the resonant Nice model with 
{imposed} migration, it is difficult to reproduce today's obliquity values, because the compactness of the initial system raises the 
frequency that tilts Saturn above the spin precession frequency of Jupiter, causing a Jupiter spin-orbit resonance crossing. Migration 
time scales sufficiently long to tilt Saturn generally suffice to tilt Jupiter more than is observed. The full $N$-body simulations 
tell a somewhat different story, with Jupiter generally being tilted as often as Saturn, but on average having a higher obliquity.
{ The main obstacle is the final orbital spacing of the giant planets, coupled with the tail of Neptune's migration.} The resonant
Nice case is barely able to simultaneously reproduce the {orbital and spin} properties of the giant planets, with a probability $\sim
0.15\%$. The loose five planet model is unable to match all our constraints (probability $<0.08\%$). The compact five planet model has
the highest chance of matching the orbital and obliquity constraints simultaneously (probability $\sim 0.3\%$).
\end{abstract}

\section{Introduction}
In the last two decades, considerable progress has been made in regards to understanding the processes that shaped the dynamical 
evolution of the Solar System, and it seems the migration of the giant planets is a critical chapter in this story. Angular momentum 
exchange with a remnant planetesimal disc induces planetary migration, moving Jupiter slightly inward and Saturn, Uranus, and Neptune 
outwards \citep{fer84,hm99}. The existence of many Kuiper belt objects in orbital mean-motion resonances (MMRs) with Neptune can be 
explained by the outward migration of Neptune. If the migration of the planets was smooth, the orbital eccentricities of the resonant 
Kuiper belt objects and the asymmetry in the resonant angle distribution of objects in the 2:1 resonance can be used to infer that 
Neptune has migrated by $\sim 7\au$ on a timescale $\ga 1\Myr$ and that the initial mass of the planetesimal disc is $\sim 50\ME$, 
{where $M_\oplus$ is the mass of the Earth} \citep{mal95,hm99,mur05}. However, this smooth migration model has difficulty reproducing 
several features of the Solar System, and alternative scenarios have been invoked.

\cite{thom99} proposed a scenario in which Uranus and Neptune grew as cores between Jupiter and Saturn, were scattered close to their 
current locations due to planet-planet gravitational interactions, and had their random velocities damped by interactions with a
remnant planetesimal disc. In this way the challenge of forming the ice giants at their current locations, with long orbital
timescales and low (estimated) solid surface density at the time of formation, could be neatly side-stepped. Their work provided the
first strong numerical evidence that the outer Solar System could have undergone considerable rearrangement from the time that the
giant planets achieved their current masses.

The {successor of this model} for the configuration of the giant planets after the protoplanetary gas disc has dissipated was presented
by \cite{tsi05} and is known as the Nice (or `classic Nice') model.  It proposes that the giant planets were once in a considerably
more compact configuration, in which Saturn started just inside the 2:1 MMR with Jupiter, and the ice giants were located at 11--13 and
13.5--$17\au$. When the 2:1 resonance was crossed by the divergent migration of Jupiter and Saturn caused by the gravitational
interactions with planetesimals, the system became dynamically unstable, exciting eccentricities and inclinations, and scattering
Uranus and Neptune to close to their present locations (in some cases, after switching their order). This scenario has many nice
properties, including possibly explaining the Late Heavy Bombardment (\citealt{gomes05}) and Jupiter's Trojan population
(\citealt{morbi05}), while typically preserving the regular satellite populations during the planetary encounters.

However, even granting the classic Nice model its successes, there were obvious questions about the history: how did the system arrive
at such an initial configuration?  The timescales for migration induced by interactions with the protoplanetary gas disc to drive the
giant planets into the inner system are shorter than the formation times for the object (e.g., \citealt{tanaka02}). The recent
revisions to the type I migration rate from a more sophisticated treatment of thermodynamics (e.g., \citealt{paard10a}) have
complicated the situation, and it remains a challenge. Building upon work done by \cite{ms01} and \cite{mc07}, a variant of the
classic Nice model was proposed by \cite{nicev2}. This {\it resonant} Nice model attempts to bridge the gap between the gas-rich phase
(typically studied with one or two planets embedded in a gas disc modelled using a hydrodynamics code) and the gas-free phase (studied
with many planets and planetesimals using an $N$-body code). The authors use the observation that convergent migration in the gas-rich
phase can result in capture into MMR and that a 3:2 MMR between Jupiter and Saturn can prevent Jupiter from migrating inwards. They
construct a number of stable configurations in which each giant planet is in resonance with its neighbours (with Saturn and the inner
ice giant in the 3:2 MMR, and the ice giants in the 4:3 or 5:4 MMR). Subsequent migration due to interactions with the planetesimals
breaks the resonances, and an instability resembling that of classic Nice occurs.

Recent studies on giant planet migration have tried to determine the actual evolution based on constraints from the secular
architecture of the giant planets \citep{mor09}, the current orbits of the terrestrial planets \citep{bra09} and
the asteroid belt \citep{mor10}. In essence, it is required that Jupiter scattered one of the ice giants outwards and thus
migrated on a very short time scale to avoid both making the orbits of the terrestrial planets too eccentric and causing secular 
resonances to sweep through the asteroid belt. Unfortunately this scenario had a very low probability of success, which led 
\cite{nes11} and \cite{bat12} to propose a scenario in which the Solar System began with five giant planets locked in MMRs. 
\cite{nes12} {have studied} a large number of five planet systems to try to reproduce the current Solar System and match all the 
constraints, which they did with a probability of about 5\% {in the best cases}.

One thing that has not yet been considered are the effects of migration on planetary obliquities (the angles between the
spin axes of the planets and their respective orbit normals) as the tilt angles of the giant planets preserve information about
migration and encounter history. The case of Saturn is particularly interesting because of a coincidence between the spin precession
rate of Saturn and the vertical secular eigenfrequency associated with Neptune, as well as the near match of Saturn's $27^\circ$ 
obliquity to that of Cassini state 2 (see Section \ref{timescale}) of the secular spin-orbit resonance. \cite{wh04} and \cite{hw04} 
{have proposed an elegant scenario}, in which the secular spin-orbit resonance is responsible for tilting Saturn (but see 
\citealt{hel09} for a potential problem with this scenario). \cite{hw04} also note that, while they concentrate on changes in secular 
eigenfrequencies due to the changing gravitational potential of a gradually decaying Kuiper belt, all that matters is that the 
frequencies change and not the underlying mechanism, and so a change in frequency due to migration should also work. At the same time, 
even though Jupiter's spin precession rate is close to the vertical secular eigenfrequency associated with Uranus \citep{war06}, {its 
$3^\circ$ obliquity means that} its obliquity evolution was very different from Saturn's.

Whether this spin-orbit resonance mechanism can work in the various migration scenarios is therefore a natural question. Here we
perform secular and particle $N$-body integrations and we initially consider {smooth migration, Nice-like, and five planet models} for 
the giant planets, and study the resulting obliquities. The obliquities of Jupiter and Saturn, in combination with the survival {and 
orbital properties} of four giant planets, provide strong constraints on the migration and encounter histories of the giant planets. 
In Section \ref{timescale} we consider the conditions needed to tilt Saturn by spin-orbit resonance. In Section \ref{numerics} we 
describe our numerical methods, and in Section \ref{section:models} our initial configurations. We present our results in Section
\ref{section:results}, with the secular and $N$-body simulations in Section \ref{section:secular} and Section \ref{section:nbody}. The
results are discussed in Section \ref{section:discussion} and summarised in Section \ref{section:disc_conclusions}.

\section{Secular spin-orbit resonance and migration timescale}
\label{timescale}
We construct a fixed reference frame with unit vectors $(\veci, \vecj,\veck)$ where $\veck$ is parallel to the total orbital angular
momentum of the system {(and hence perpendicular to the invariable plane) and $\veci$ and $\vecj$ are in the invariable plane.}
Consider a planet whose unit vector of the spin direction $\vecs$ is tilted at an angle $\varepsilon$ from the unit orbit normal
$\vecn$, i.e., $\cos \varepsilon = \vecs \cdot \vecn$. The orbit has an inclination $\cos i = \vecn \cdot \veck$. If there are no
perturbations on the orbit and thus $\vecn$ is fixed in space, the torque from the Sun on the oblate figure of the planet (and on any
satellites whose orbits are locked to the planet's equator plane) causes $\vecs$ to precess uniformly about $\vecn$ at the frequency
$\alpha (1-e^2)^{-3/2} \cos\varepsilon$. Here $e$ is the orbital eccentricity and $\alpha$ is the precession constant, given by
\begin{equation}
\alpha = \frac{3 G M_\odot}{2 \nu a^3}\left(\frac{J_2 + q'}{\mathcal{C} + \ell}\right) ,
\end{equation}
{where $G$ is the gravitational constant, $M_\odot$ is the solar mass, $a$ is the orbital semi-major axis of the planet, $\nu$ is
its rotation frequency, $J_2$ is the second zonal harmonic coefficient of its gravity field, $\mathcal{C}$ is its moment of inertia
normalised to $m R^2$, $m$ and $R$ are the mass and radius of the planet, $\ell$ is the orbital angular momentum (normalised to $m R^2
\nu$) of the satellites whose orbits are locked to the planet's equator, and $q'/J_2$ is the ratio of the torque on the satellites
to that directly exerted on the planet (see \citealt{wh04} for explicit definitions of $\ell$ and $q'$).}

Secular perturbations from other planets force $\vecn$ to precess about $\veck$. In the simple case where $\vecn$ precesses at a
constant rate $s$, there are either two or four locations where $\vecs$ remains coplanar with $\vecn$ and $\veck$, and $\vecs$ and
$\vecn$ precess at the same rate $s$ about $\veck$. These secular spin-orbit resonance configurations are called the Cassini states
\citep{col66,pea69,pea74}.

\cite{wh04} and \cite{hw04} suggest that Saturn's obliquity of nearly 27$^\circ$ may be the result of Saturn being caught in Cassini
state 2 with the vertical secular eigenfrequency $s_8$ associated with Neptune.\footnote{{We follow \cite{bre74} and \cite{las90}
and denote the vertical secular eigenfrequencies associated with the giant planets by $s_6$, $s_7$, and $s_8$. \cite{wh04} and
\cite{hw04} use the notation $g_{16}$, $g_{17}$, and $g_{18}$; \cite{app86} use $g_6$, $g_7$, and $g_8$; while \cite{md} use $f_6$,
$f_7$, and $f_8$.}} {They} show that Saturn's obliquity $\epsS$ can be increased from an initially small value to 27$^\circ$ by capture
into Cassini state 2 if $|s_8/\alpha_{\rm S}| > 1$ initially and decreases to $<1$. They also show that the more complicated nodal
regression of Saturn caused by perturbations from Jupiter and Uranus do not affect the capture into spin-orbit resonance and the
associated growth in Saturn's obliquity.

One way to decrease $|s_8|$ (and {hence} $|s_8/\alpha_{\rm S}|$) is through the outward migration of Neptune. The time scale $\tau_s =
\alpha_{\rm S}/{\dot s}_8$ for the decrease in $|s_8|$ must be adequately long for Saturn's spin to stay in resonance.
{The critical time scale $\tau_{s,{\rm min}}$ can be estimated from requiring the time that it takes for the obliquity of Cassini state
2 to move by an angle comparable to the width of the separatrix to be longer than the libration period about Cassini state 2. With the
obliquity of Cassini state 2 given by $\cos\epsS \approx-(s_8/\alpha_{\rm S}) \cos i$ (when $\epsS$ is large), the half width
of the separatrix $\Delta \epsS = 2 (\tan i/\tan \epsS)^{1/2}$, and the libration frequency $\omega_{\rm lib} = (-\alpha_{\rm S}
s_8\sin\epsS \sin i)^{1/2}$ for small-amplitude libration about Cassini state 2 \citep{hen87,wh04,hw04}. Then
$2 \Delta \epsS/{\dot \varepsilon}_{\rm S} \approx 2\pi/\omega_{\rm lib}$ gives

\begin{equation}
\tau_{s,{\rm min}} \approx \frac{\pi}{2\alpha_{\rm S}\sin i\sin\epsS} .
\label{eq:tsmin}
\end{equation}
The condition ${\dot s}_8 \approx \omega_{\rm lib}^2$ found by \cite{hw04} gives a similar $\tau_{s,{\rm min}}$ (but without the
coefficient $\pi/2$). \cite{bou09} have also shown using a different analytic argument that the critical time scale $\tau_{s,{\rm min}}
\propto 1/(\alpha_{\rm S} i)$.}

\begin{figure}[t]
\epsscale{1.1}
\plottwo{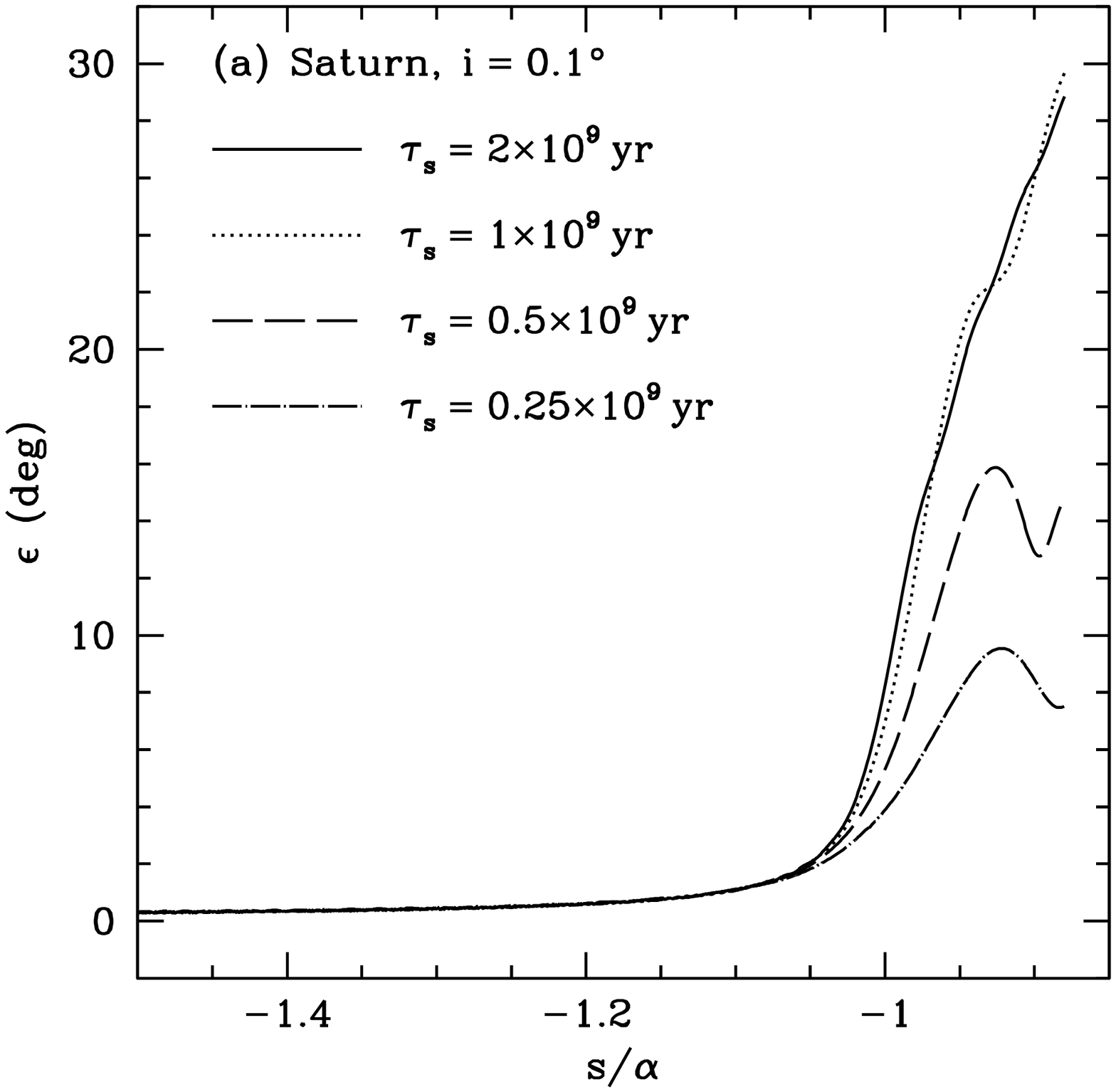}{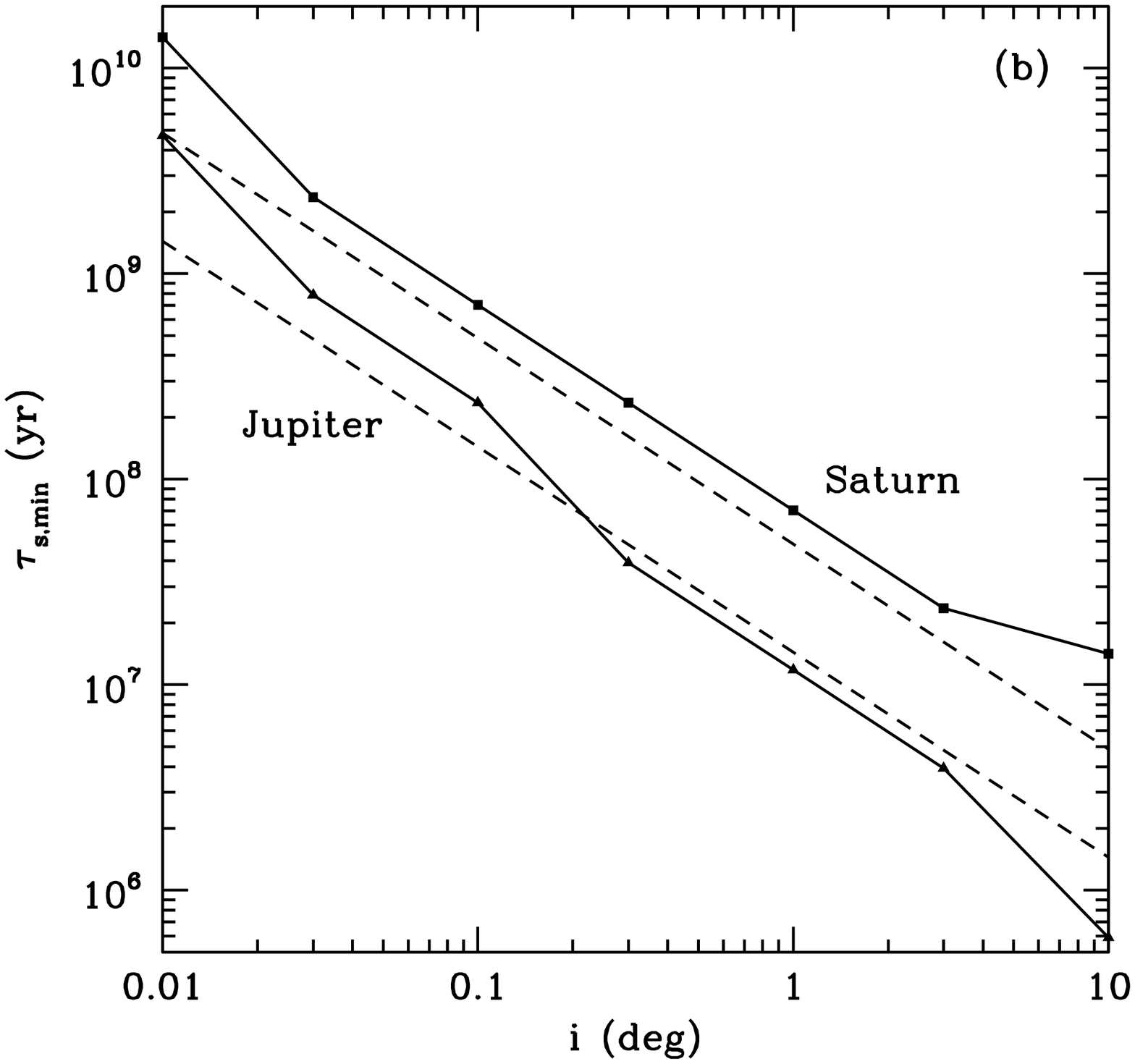}
\caption{(a) Evolution of Saturn's obliquity $\varepsilon$ in a set of simulations in which Saturn's orbit normal $\vecn$ is inclined
by $i=0.1^\circ$ from the normal to the invariable plane $\veck$ and forced to precess about $\veck$ at rate $s$. The
orbital eccentricity $e = 10^{-3}$, the spin precession constant $\alpha_{\rm S} = 0.79\arcsec\,{\rm yr}^{-1}$, and $s$ changes
linearly from $-5\alpha_{\rm S}$ to $-0.88\alpha_{\rm S}$ according to $s(t) = (-5 + t/\tau_g) \alpha_{\rm S}$. Saturn remains in
Cassini state 2 to the end of the simulation if $\tau_s > \tau_{s,{\rm min}} \approx 700$~Myr or escapes from the resonance before the
end of the simulation if $\tau_s < \tau_{s,{\rm min}}$. (b) Critical time scale $\tau_{s,{\rm min}}$ as a function of orbital
inclination $i$ for both Saturn and Jupiter {(with $\alpha_{\rm J}=2.67\arcsec\,{\rm yr}^{-1}$). The dashed lines show
$\tau_{s,{\rm min}}$ according to Equation (\ref{eq:tsmin}) with the final obliquity $\varepsilon \approx 29^\circ$.}
\label{fig:saturn_only}}
\end{figure}

We have verified {Equation (\ref{eq:tsmin})} with numerical simulations using the symplectic integrator SyMBA, which was modified to
include the evolution of the spin axis (\citealt{lee07}; see also Section 3). We integrated Saturn alone with $\alpha_{\rm S} =
0.79\arcsec\,{\rm yr}^{-1}$ \citep{tre91}, $e = 10^{-3}$ and imposed orbital nodal precession rate $s$ that changes linearly from
$-5\alpha_{\rm S}$ to $-0.88\alpha_{\rm S}$ according to $s(t) = (-5 + t/\tau_s) \alpha_{\rm S}$. The spin axis is initially parallel
to $\veck$. Fig.~\ref{fig:saturn_only}(a) shows the evolution of Saturn's obliquity in a set of simulations with $i = 0.1^\circ$ and
different $\tau_s$. Saturn is captured into Cassini state 2 and remains there until the end of the simulation (when $\epsS$ is slightly
larger than $27^\circ$) if $\tau_s > \tau_{s,{\rm min}} \approx 700$~Myr. For $\tau_s < \tau_{s,{\rm min}}$, Saturn is initially
captured into Cassini state 2, but it escapes from the resonance when the obliquity becomes too large; the final obliquity increases
with increasing $\tau_s$ and can be more than $10^\circ$ for $\tau_s \ga \frac{1}{2}\tau_{s,{\rm min}}$. From similar simulations
with $i = 0.01^\circ$--$10^\circ$, we have determined $\tau_{s,{\rm min}}$ as a function of $i$, and the results are shown in
Fig.~\ref{fig:saturn_only}(b) for Saturn, {as well as Jupiter with $\alpha_{\rm J}=2.67\arcsec\,{\rm yr}^{-1}$}. We confirm
{Equation (\ref{eq:tsmin})} over a wide range in $i$ and find that for Saturn $\tau_{s,{\rm min}} \sim 20 \Myr$ for $i$ of a few
degrees. We note that $20\Myr$ is roughly the upper limit of the typical migration time scale of the giant planets caused by
planetesimal scattering \citep{mor10}. Also, in our earlier simulations of the classic Nice model (\citealt{lee07}), we found that the
orbital inclination of Saturn can reach $\sim 5^\circ$ (and those of the ice giants $10$--$15^\circ$) during the
encounter phase in some simulations (see, e.g., Fig.~2 of \citealt{lee07}).

Summarising, it appears possible to tilt Saturn to its current obliquity if $|s_8/\alpha_{\rm S}|$ changes slowly enough. In the next
sections we investigate whether or not this can be realised {in the various migration scenarios}.

\section{Numerical methods}
\label{numerics}
To study the obliquity evolution in the various migration scenarios, we performed several types of numerical simulations of the Sun and
the giant planets. One set of simulations consisted of integrating the secular evolution of the migrating giant planets without
encounters and mean-motion resonances. The second set of simulations integrated the full $N$-body problem {with planetesimals}. These
integrations used the Kepler-adapted symplectic $N$-body code SyMBA \citep{dll98}, a descendant of the original techniques of
\cite{wis91} and \cite{kyn}, as modified by \cite{lee07} to incorporate spin evolution. {There are some additional modifications for
both the secular and $N$-body simulations that we now describe.}

\subsection{Secular Simulations}
\label{secspin_descr}
Since we expect the obliquity evolution to be driven by resonances between the nodal regression and spin precession frequencies, both
rather small compared to the orbital frequencies, a useful approximation to the full $N$-body behaviour for exploratory simulations can
be obtained by modelling the purely secular evolution of the system. We have adapted the spin-SyMBA code of \cite{lee07} to remove
mean-motion effects and close encounters when desired. The method is the following.

We consider a system made up of the Sun and some number of planets. The planets are assumed to undergo principle axis rotation, and the
effects of the oblateness of the planets on the orbital evolution are neglected. The integration step has the form
\begin{equation}
M^{\,\tau/2} \, R^{\,\tau/2} \, S^{\,\tau/2} \, D^{\,\tau} \, S^{\,\tau/2} \, R^{\,\tau/2} \, M^{\,\tau/2}
\label{eq:step}
\end{equation}
where $\tau$ is the time step and each term is an operator. The operator $D$ advances the planets along their osculating Kepler
orbits, $S$ advances the spin axes according to the mutual torques between planets and the torques due to the Sun, $R$ handles the
secular interactions between the planets, and $M$ generates radial migration (whose recipe is discussed {later in this subsection}).
Unlike with the full code described by \cite{lee07}, the integration of the secular system is performed in heliocentric coordinates
rather than canonical heliocentric, and so there is no linear drift term to correct for the barycentric momentum of the Sun
\citep{dll98}. Since we do not treat close encounters in the secular approximation, we do not need the recursive subdivision of the
time step of SyMBA. The Kepler drift operator is straightforward, and the spin torque operator follows \cite{lee07}, but the secular
operator replacing the interaction term is new.

The secular interaction between the planets can be treated in several ways. Here we forgo the eigenvalue approach and instead directly
integrate Lagrange's planetary equations with the first-order secular potential (see, e.g., \citealt{md}). Let $a$, $e$, $i$ be the
semi-major axis, eccentricity, and inclination of a planet, $\varpi$ the longitude of perihelion, $\Omega$ the longitude of the
ascending node, and $n$ the mean motion. To apply the $R$ operator, for each pair of planets (with indices $j, k; j \neq k$) we write
the disturbing function on planet $j$ due to $k$ as
\begin{equation}
R_{jk} = n_j \, a_j^2 \, 
\left[\frac{1}{2} \, A_{jj} \, e_j^2 + A_{jk} \, e_j \, e_k \,\cos(\varpi_j -\varpi_k) +
 \frac{1}{2} \, B_{jj} \, i_j^2 + B_{jk} \, i_j \, i_k \,\cos(\Omega_j
 -\Omega_k)\right] .
\end{equation}
Here the matrix elements $A$ and $B$ are functions of the masses and semi-major axes and they contain information about the coupling
between the planets \citep{bro50}. We can then compute the time derivatives
\begin{align}
{\dot e_{jk}} & = -\frac{1}{n_j \, a_j^2 \, e_j} \frac{\partial R_{jk}}{\partial \varpi_j}, \,\, \quad
{\dot i_{jk}} =  -\frac{1}{n_j \, a_j^2 \, i_j} \frac{\partial R_{jk}}{\partial \Omega_j} , \\
{\dot \varpi_{jk}}  & =  \frac{1}{n_j \, a_j^2 \, e_j} \frac{\partial R_{jk}}{\partial e_j}, \,\,\quad
{\dot \Omega_{jk}} = \frac{1}{n_j \, a_j^2 \, i_j} \frac{\partial R_{jk}}{\partial i_j} .
\end{align}
The total rate of change in eccentricity of planet $j$ is given by
\begin{equation}
{\dot e_{j}} = \sum_{k \neq j} {\dot e_{jk}}
\end{equation}
and similarly for the changes in the other elements ${\dot i_j}, {\dot \varpi_j},$ and ${\dot \Omega_j}$. These equations are
integrated using a fourth-order Runge-Kutta scheme.

It may seem odd to evolve what is fundamentally a secular ring problem using point particles, as this requires repeatedly solving
Kepler's equation during the Kepler drifts, which should be unnecessary in a secular model. However, in this way the well-tested spin
vector infrastructure of the \cite{lee07} modifications to the SyMBA code could be reused. We have confirmed by comparing with a
different secular code (one which solves the corresponding eigenproblem instead) and with the full $N$-body integration that this
approach recovers the expected evolution.

Note that the secular step itself is rather time-consuming compared to a simple force evaluation due to the need to compute multiple
trigonometric functions and the Laplace coefficients. The separation between the orbital time scale and the secular, spin, and
migration time scales provides several opportunities for optimisation. Advancing the secular term $R$ every (orbital time scale) step
is unnecessary, and applying it every 10 steps serves to make the runtime comparable to the $N$-body run, without changing the
resulting evolution. Another improvement can be obtained by dividing each of the secular, spin, and migration time scales by some
constant factor (making sure to keep them several orders of magnitude longer than the orbital time scale) during the simulation and
then scaling the resulting output. This is possible because the behaviour of the spin-orbit resonance depends not on the absolute
values of these time scales but on their ratios. In our regime, using a factor of 10--20 produces negligible errors relative to the
original simulation and provides almost all of the possible speed benefits. We have verified that our results are insensitive to these
approximations, which are used only in the secular integrations.

In the standard picture, the final configuration of the giant planets should evolve mostly due to mutual interactions and interactions
with an external remnant planetesimal disc (such as the Kuiper belt). Accordingly, with the possible exception of objects which get
scattered out into the Kuiper belt and Scattered Disk \citep{1997Sci...276.1670D}, imposing a prescription for migration and random
velocity damping during the secular simulations should be a reasonable approximation to the effect of the planetesimal disc.
Let $t$ be the time since the start of the simulation, and $\tau$ the migration time scale. We follow \cite{lee07}, and take as
migration expressions
\begin{equation}
{\dot a}_k = \frac{\Delta a_k}{\tau} \exp(-t/\tau) ,
\label{eq:dakdt}
\end{equation}
and
\begin{equation}
\quad {\dot a}_k = \frac{\Delta a_k}{\tau}
 \left(\frac{t}{\tau}\right) \exp[-t^2/(2\tau^2)] ,
\label{eq:dakdtice}
\end{equation}
where $\Delta a_k$ is a parameter measuring the amount of distance the planet should travel due to the migration to end up on its
current orbit. {In the secular problem, $\Delta a_k$ is simply the difference between the current value and the starting value.}
Equation~(\ref{eq:dakdt}) results in faster migration at earlier times than Equation~(\ref{eq:dakdtice}), but the two resulting 
trajectories meet again at $t=2 \tau$ when {the distance travelled is} $\sim\!0.86 \Delta a_k$, after which 
Equation~(\ref{eq:dakdtice}) has faster migration. Equation~(\ref{eq:dakdt}) has a continually decreasing ${\dot a}$, whereas 
Equation~(\ref{eq:dakdtice}) has ${\dot a}$ increasing until a maximum at $\tau$ (corresponding to a maximum speed of $\sim\!0.6 
\Delta a_k/\tau$) beyond which it decays to zero. {The $M$ operator of Equation (\ref{eq:step}) simply applies the changes in $a$ to 
the planets. As in \cite{lee07}, we evolve all planets under Equation (\ref{eq:dakdt}) in our integrations of the smooth migration 
model, and we use Equation (\ref{eq:dakdt}) for Jupiter and Saturn and Equation (\ref{eq:dakdtice}) for Uranus and Neptune in our 
integrations of the Nice model. Note that we do not apply any eccentricity or inclination damping to the secular simulations: as there 
is no stirring mechanism, $e$ and $i$ would damp too quickly to serve as a useful model for the spin-orbit resonance crossing. We 
evolve every simulation to $10 \tau$.}

This concludes our description of the secular simulations. We now move on to discuss the full $N$-body simulations.

\subsection{\mbox{\boldmath $N$}-body simulations}
\label{sect:mig}
In the full model the planets migrate because of their interaction with the planetesimal disc and by mutual encounters. We integrated
the giant planets and a planetesimal disc consisting of several thousand planetesimals using the spin-SyMBA code from \cite{lee07}. We
made an additional modification to the code so that it only outputs the orbital elements of the planets and not those of the
planetesimals in the disc, and stops the integration once the number of planets is fewer than four. The initial conditions for these
simulations are described in the next section and are similar to those of \cite{nes12}. The outer Solar System and the obliquities of
the giant planets were integrated with a time step of 0.35~yr. Planets and planetesimals were removed once they were further than
1000~AU from the Sun (whether bound or unbound) or when they collided with a planet or ventured closer than 0.3~AU to the Sun. We
evolve every simulation to 500~Myr or until fewer than four planets remain, whichever occurs first. The simulations were run on the 
TIARA computer cluster at the Institute for Astronomy and Astrophysics, Academia Sinica, in Taiwan.

For these $N$-body simulations, a minor adjustment to the SyMBA algorithm as implemented in the Swift package\footnote{Available at
http://www.boulder.swri.edu/\~{}hal/swift.html} is needed. SyMBA decomposes the Hamiltonian in canonical heliocentric coordinates
i.e. heliocentric positions and barycentric velocities, and employs a multiple time step technique to handle close encounters.
When we first integrated our initial conditions for the {resonant Nice} configurations in the absence of {planetesimals}, we
found unexpectedly that no more than half of the initial conditions were stable for $1\,$Gyr when we used SyMBA, whereas they were all
stable under both the original Wisdom-Holman method and a pure canonical heliocentric integrator i.e. without any treatment of
close encounters. This instability is caused by Jupiter and Saturn being in the 3:2 resonance, where they are close enough at
conjunctions that they are considered to be undergoing encounters by the SyMBA algorithm, and their time steps are subdivided at every
conjunction.
The time step subdivision is achieved in the SyMBA algorithm by using a partition function to decompose the $r^{-2}$ gravitational
force between two planets into forces that are non-zero only between two cut-off radii, $R_{k+1} \le r < R_{k}$. The simplest partition
function is the ($2\ell+1$)th order polynomial in $x$ that has $f_\ell(0) = 1$, $f_\ell(1) = 0$, and all derivatives up to the $\ell$th
derivatives zero at $x = 0$ and $1$. \cite{dll98} found that the third-order polynomial $f_1(x) = 2 x^3 - 3 x^2 + 1$ worked well for
many situations, which did not include repeated encounters on orbital time scales over $100\,$Myr--$1\,$Gyr. For this more challenging
situation, a smoother partition function is needed, and the next appropriate polynomial is 
$f_3(x) = 20 x^7 - 70 x^6 + 84 x^5 - 35 x^4+ 1$. We found that the use of $f_3(x)$ sufficed to keep each of the resonant Nice initial
conditions stable under SyMBA for $1\,$Gyr. Of course, in our actual simulations with {planetesimals}, the planets will have migrated
on time scales much shorter than the numerical instability time scales, but it is useful for testing purposes that the integrator does
not fail on the time scales of interest.

\section{Models and initial conditions}
\label{section:models}
We restrict ourselves to four classes of initial conditions, namely a version of the smooth migration scenario of \cite{hm99} with
four planets, a variant of the resonant Nice model \citep{nicev2} with four planets, and two five planet models from \cite{nes12}. We
integrate the \cite{hm99} initial conditions with the secular equations only, the resonant Nice with the secular equations and full
$N$-body model, and the five planet cases with full $N$-body only.

\subsection{Smooth Migration}
We set the initial semi-major axes of Jupiter and Saturn to be $a_{\rm J} = 5.40 \au$ and $a_{\rm S} = 8.73 \au$, respectively. Uranus
and Neptune are given random semi-major axes uniformly drawn from the ranges $14.19$--$18.19 \au$ and $21.07$--$25.07 \au$. This
configuration means that Saturn begins just beyond the 2:1 resonance with Jupiter at $8.6 \au$. Eccentricities are set to 0.001.

\subsection{Resonant Nice Configuration}
In this case, the four giant planets are all initially in resonance. Saturn and Jupiter are in the 3:2, as are Saturn and 
the inner ice giant. There are two distinct stable configurations for Uranus and Neptune found by \cite{nicev2}, namely the 4:3 and 
the 5:4; we study only the 4:3, {which has} the outer ice giant at 11.6~AU.
To generate the initial conditions, we used an approach similar to that of \cite{nicev2}, but instead of a hydrodynamic code,
we employed the $N$-body code SyMBA, with forced migration and eccentricity damping to model the effects of the proto-planetary gas
disc. We began with Jupiter and Saturn and placed them just outside the 3:2 mean-motion resonance. Jupiter was forced to migrate
outwards and Saturn inwards, and the migration rates were chosen so that they would be nearly stationary after capture into the 3:2
resonance. {The eccentricity damping rate is specified by \citep{lee07}}
\begin{equation}
{\dot e}_k/e_k = -K_e |{\dot a}_k/a_k| .
\label{eq:ei}
\end{equation}
We adopted $K_e = 100$ for Jupiter and adjusted $K_e$ for Saturn to match the equilibrium eccentricities reported by \cite{nicev2}.
After the Jupiter-Saturn pair had reached equilibrium in the 3:2 resonance, we added an ice giant just outside the 3:2 resonance with 
Saturn and forced it to migrate inwards until Saturn and the ice giant reached equilibrium in the 3:2 resonance. This process was
repeated for the 4:3 resonance between the ice giants. For the ice giants, which should be undergoing type I migration, we assumed 
that the values of $K_e$ are identical and adjusted $K_e$ until we matched the eccentricities reported by \cite{nicev2} at each stage 
of this process. To model the dispersal of the gas disc, we continued the integration with an exponential decay in the migration and 
eccentricity damping rates.

\subsection{Five-Planet Configurations}
For the five planet cases we have chosen two configurations which bracket the most compact and least compact configurations studied by
\cite{nes12}. The configurations we studied were {3:2, 3:2, 4:3, 4:3} and 3:2, 3:2, 2:1, 3:2. In the first case the outermost ice giant
was at 14.2~AU while in the latter case it was at 22.2~AU. The initial conditions for the first (compact) configuration were generated
in a similar manner to the resonant Nice configuration, {but with the addition of a third ice giant in 4:3 resonance with the
second one}. The initial conditions for the second (loose) configuration were provided by D. Nesvorn\'{y}. These two configurations
bracket the resonant Nice model of \cite{nicev2} and the model of \cite{hm99} in terms of spacing between the planets, {but with an
additional ice giant.}

For the initial conditions of the $N$-body simulations {with planetesimals}, we selected randomly from the last five frames of output
from the gas dispersal integration for the {resonant Nice and compact five planet cases}, and the system is rescaled so that initially
Jupiter's semi-major axis $a_{\rm J} = 5.45\au$. We confirmed that the last five frames are stable for 1~Gyr (in the absence of forced
migration and damping), after we made the reported minor adjustment to the SyMBA algorithm (see Section \ref{sect:mig}).
For the initial conditions provided by D. Nesvorn\'{y} we generated five frames by integrating the planetary system without the
planetesimal disc for 500~yr and output the orbital elements every 100~yr.

The planetesimal disc consisted of 2000 planetesimals for the resonant Nice and compact five planet cases and 3000 for the loose
five planet case. The surface density of the planetesimal disc scales with heliocentric distance as $\Sigma \propto r^{-1}$. The outer
edge of the disc was always set at 30~AU. The inner edge was $\Delta = 0.5$~AU from the outermost ice giant in the resonant Nice and
compact five planet cases and $\Delta = 1$~AU for the loose five planet case. The total disc mass was 50$M_{\oplus}$ for the resonant
Nice model, 35$M_{\oplus}$ for the compact five planet case and 20$M_{\oplus}$ for the loose five planet case. \cite{nes12} have found
that a disc of $\sim 50M_\oplus$ gives the best results for the specific resonant Nice configuration that we adopted. We decreased the
disc mass for the compact five planet case to account for the mass of the extra ice giant. All configurations have roughly the same 
surface density in planetesimals. We generated 256 different realisations of each frame by giving a random deviation of $10^{-6}$~AU 
to {the semi-major axis of} each planetesimal, for a total of 1\,280 simulations per system.

In all cases, we used the current masses and spin precession constants of the giant planets, with the latter taken from \cite{tre91}.
For the current {orbital semi-major axes}, the spin precession constants are $\alpha_{\rm J} = 2.67\arcsec\,{\rm yr}^{-1}$,
$\alpha_{\rm S} = 0.793\arcsec\,{\rm yr}^{-1}$, $\alpha_{\rm U} = 0.0448\arcsec\,{\rm yr}^{-1}$, and $\alpha_{\rm N} =
0.0117\arcsec\,{\rm yr}^{-1}$ for Jupiter, Saturn, Uranus, and Neptune, respectively. There is an ambiguity about interpreting the ice
giants' values {in the resonant Nice and five planet cases, because the ice giants can switch places and, in the five planet cases, any
one of the ice giants can be ejected.} However, the masses {of Uranus and Neptune} are very similar (14.5$M_\oplus$ and
17.1$M_{\oplus}$), and in an odd coincidence, their spin precession constants are also very similar when scaled to a common semi-major
axis, so the ambiguity presents much less difficulty in practice than might have been expected. Thus we adopted the parameters of
Uranus and Neptune for the initially inner and outer ice giants in the resonant Nice model, {and added a third ice giant with a mass
of $15.8 M_\oplus$ (the average mass of Uranus and Neptune) and the precession constant of Neptune for the compact five planet
configuration. For the loose five planet configuration provided by D. Nesvorn\'{y}, the masses of the initially inner, middle, and 
outer ice giants were equal to $15 M_{\oplus}$, the mass of Uranus, and the mass of Neptune, respectively, while the precession 
constants were the same as in the compact case.} We set the initial spin vectors of the planets to be only slightly perturbed (0.001 
radians) from the inertial $z$-axis. The initial orbital inclinations are specified below for the different types of simulations. All 
other angles are set at random values.

\section{Results}
\label{section:results}
In this section we present the results of our {secular} and $N$-body simulations. We discuss the smooth migration case first to explain
the underlying mechanism of how to tilt the gas giants, followed by the resonant Nice and five planet simulations.

\subsection{Secular Simulations}
\label{section:secular}
To understand how the secular spin-orbit resonance scenario works in the full planetary migration context, but without the complicating
factors present in noisy $N$-body simulations, we apply the secular spin approach of Section~\ref{secspin_descr} as an intermediate
scheme between the one-planet simulations in Section~\ref{timescale} and the full $N$-body runs to {two} of the histories
{(smooth migration and resonant Nice)} described in Section \ref{section:models}.

\subsubsection{Smooth migration}
\begin{figure}
\epsscale{1.0}
\plotone{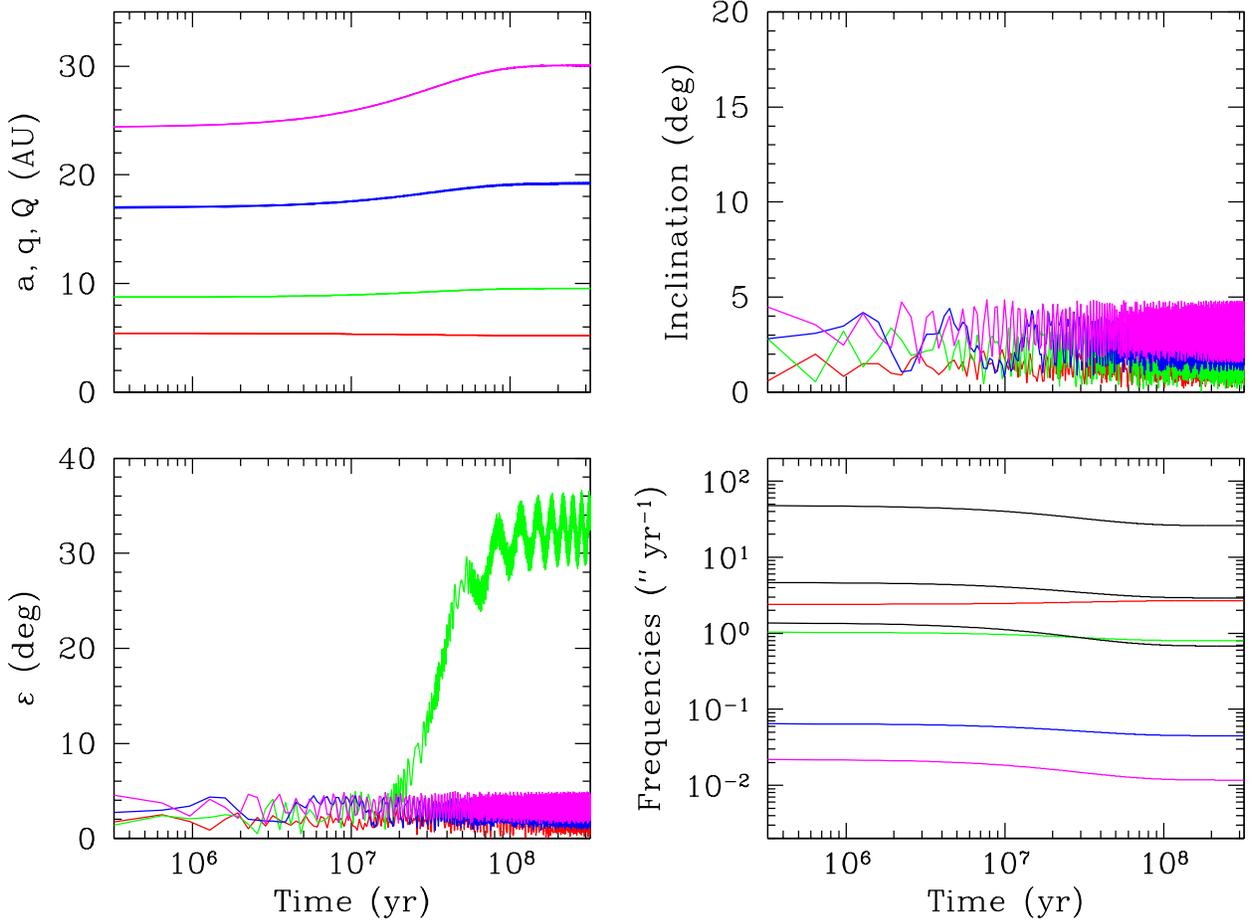}
\caption{Evolution of a successful purely secular smooth migration model with migration timescale $\tau=32\Myr$ {and initial
inclinations $i = 2^\circ$}. The upper left panel shows the orbital semimajor axis $a$, perihelion distance $q$, and aphelion distance
$Q$ for Jupiter (red), Saturn (green), Uranus (blue), and Neptune (magenta).  (Note that here the eccentricities are low enough that 
the lines cannot be distinguished.)  The upper right and lower left panels show the orbital inclinations $i$ and the obliquities
$\varepsilon$, respectively. In the lower right panel, the coloured lines correspond to the spin precession frequencies $\alpha$ of the
planets, and the three black lines are the nondegenerate vertical secular eigenfrequencies $|s_6|$ (highest), $|s_7|$ (middle), and
$|s_8|$ (lowest). Note that $|s_8|$ is initially between the spin precession frequencies of Jupiter and Saturn.}
\label{fig:smooth_good} 
\end{figure}
\begin{figure}
\plotone{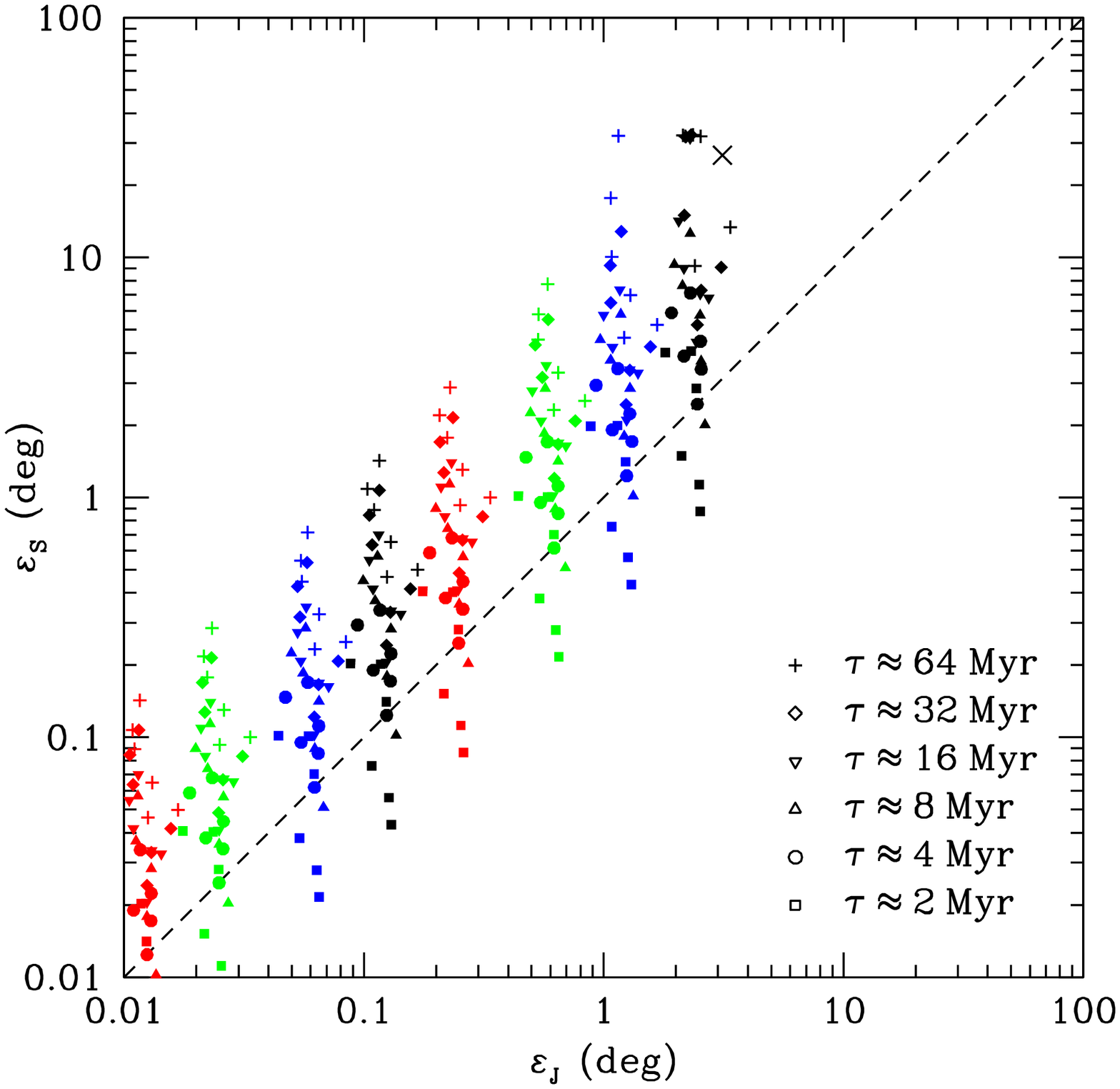}
\caption{Resulting obliquities for Jupiter ($\epsJ$) and Saturn ($\epsS$) in secular simulations of the smooth migration model as a
function of $\tau$. Each colour corresponds to a different initial inclination, from 0.01${^\circ}$ on the left to $2^{\circ}$ on
the right.  The diagonal cross marks the observed configuration of Jupiter and Saturn.}
\label{fig:smooth_JS_overview} 
\end{figure}
The simplest model to consider is a purely secular model of the smooth migration scenario, as it behaves as the analytic theory would
predict. Figure~\ref{fig:smooth_good} presents an overview of a
successful run that reproduces the current obliquities of the gas giants. In this example $\tau=32\Myr$ {and initial $i = 2^\circ$ for
all planets}. The upper panels describe the evolution of the orbits, with the migration in semi-major axes imposed by
Eq.~(\ref{eq:dakdt}). The inclinations oscillate around the forced values due to secular interactions. The lower panels describe the
evolution of the system spins. The obliquities $\varepsilon$ of all four planets remain low until approximately $20\Myr$, during which
time the evolution of $\varepsilon$ is driven entirely by changes in the inclination, as can be seen by comparing the bottom left and
top right panels. However, after about $20\Myr$ the obliquity of Saturn rises to a mean of $\sim\!33^{\circ}$ with a superimposed
oscillation of about $4^{\circ}$. That this obliquity rise is due to capture into a secular spin-orbit resonance is clear from the
bottom right panel, which compares the spin precession frequencies $\alpha$ of the planets with the three non-degenerate vertical
secular eigenfrequencies $|s_6|$, $|s_7|$, and $|s_8|$ of the system. At the beginning of the simulation, no eigenfrequency is
particularly close to any spin precession frequency of a planet, with the slowest eigenfrequency $|s_8|$ (which is that dominated by
Neptune's contribution) situated between the spin precession frequencies of Jupiter and Saturn. 
However, as the semi-major axes change due to migration, so do both the eigenfrequencies and the spin precession frequencies. After
$20\Myr$, when Saturn is observed to begin tilting over, we see that $|s_8|$ matches the spin precession frequency of Saturn, and it is
this secular spin-orbit resonance which could explain the current tilt.

Not every set of parameters results in such a Saturnian obliquity: as discussed in Section~\ref{timescale}, both $\tau$ and $i$ play a
role such that $\tau \times i$ {must exceed a certain value}. Fig.~\ref{fig:smooth_JS_overview} shows the resulting Jovian and
Saturnian obliquities from simulations with $\tau = 1$--$100\Myr$ and initial $i = 0.01$--$2^\circ$.  In general, as the time scale
increases so does the resulting Saturnian obliquity, up to a maximum of $\sim\!30^{\circ}$. This is very close to the observed
$27^{\circ}$, thus providing a satisfying explanation for Saturn's obliquity. Moreover, in this migration time scale regime, at least
an inclination of $1$--$2^{\circ}$ is required, which also gives a Jovian obliquity compatible with the current value. Thus smooth
migration can work at large $\tau \times i$ ($\ga$ 30~Myr\,deg): the minimum $\tau \times i$ producing a simulation with final
Saturnian obliquity $>15^{\circ}$ is 32~Myr\,deg, although some simulations with $\tau \times i$ above this limit have final Saturnian
obliquity $<15^{\circ}$. This joint constraint will prove to be useful.

Although the smooth migration model is able to reproduce Saturn's obliquity if $\tau \times i$ is sufficiently large,
we want to emphasise that it has great difficulties in explaining the current secular architecture of the giant planets \citep{mor09}
and the terrestrial planets \citep{bra09}, as well as the orbital properties of the main belt asteroids \citep{mor10}, and should not
be considered as a proxy for the past evolution of the Solar System.

\subsubsection{Resonant Nice}
\label{section:secular_nice}
In principle the situation with the resonant Nice model should be very similar. The outward migration of Saturn and the ice giants
will lower the spin precession frequency of Saturn slightly and the eigenfrequencies considerably, so the same resonance crossing
should occur, and it does. The difficulty becomes apparent when we examine an equivalent to the smooth migration run shown in
Fig.~\ref{fig:smooth_good}.

\begin{figure}
\plotone{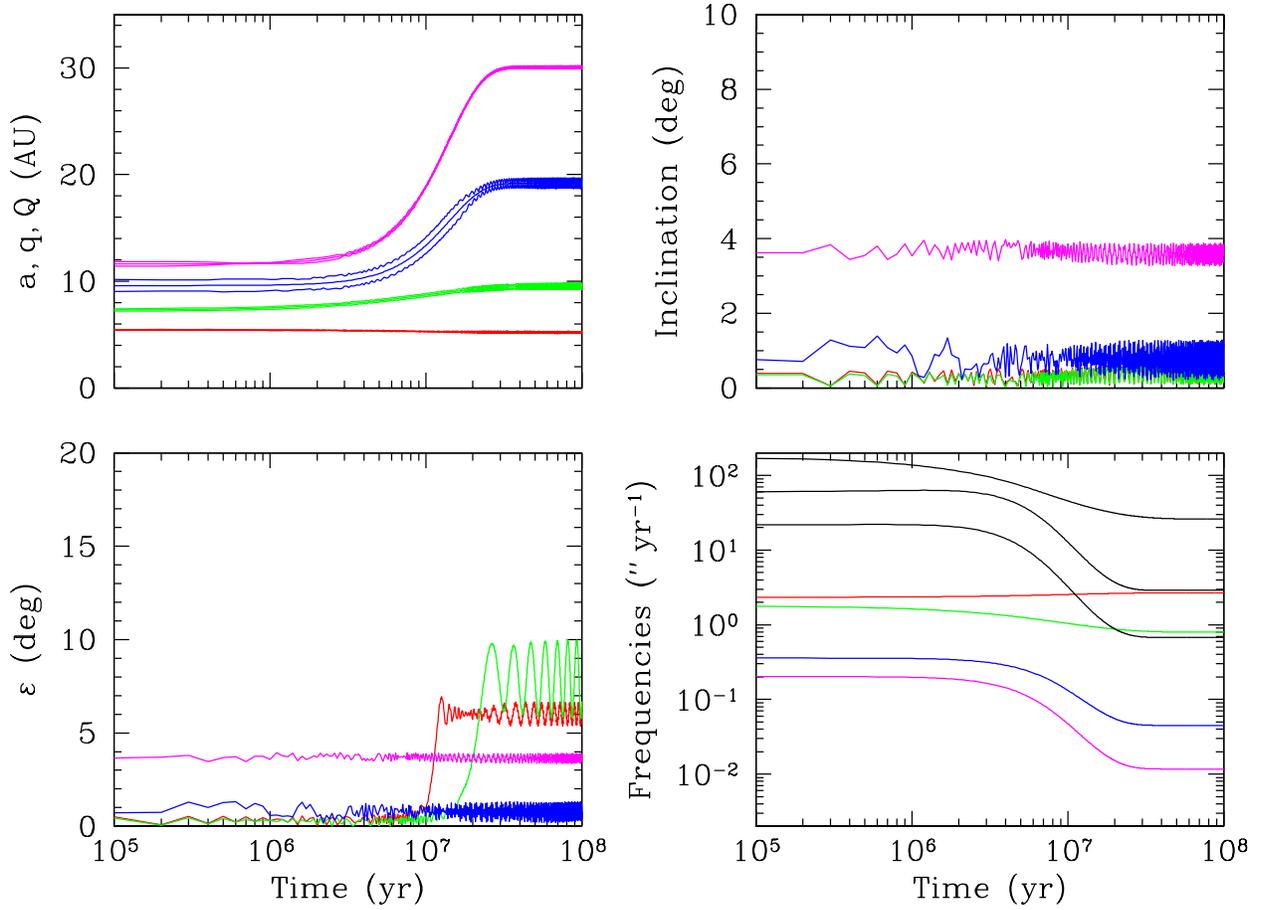}
\caption{Same as Fig. \ref{fig:smooth_good}, but for a secular resonant Nice model with $\tau= 10\Myr$ and $i_{\mathrm{N}} =
4^{\circ}$. Note that $|s_8|$ now crosses the spin precession frequency of Jupiter before it reaches that of Saturn, resulting in
Jupiter tilting to $\simeq 6^{\circ}$.}
\label{fig:secular_nice} 
\end{figure}
\begin{figure}
\plotone{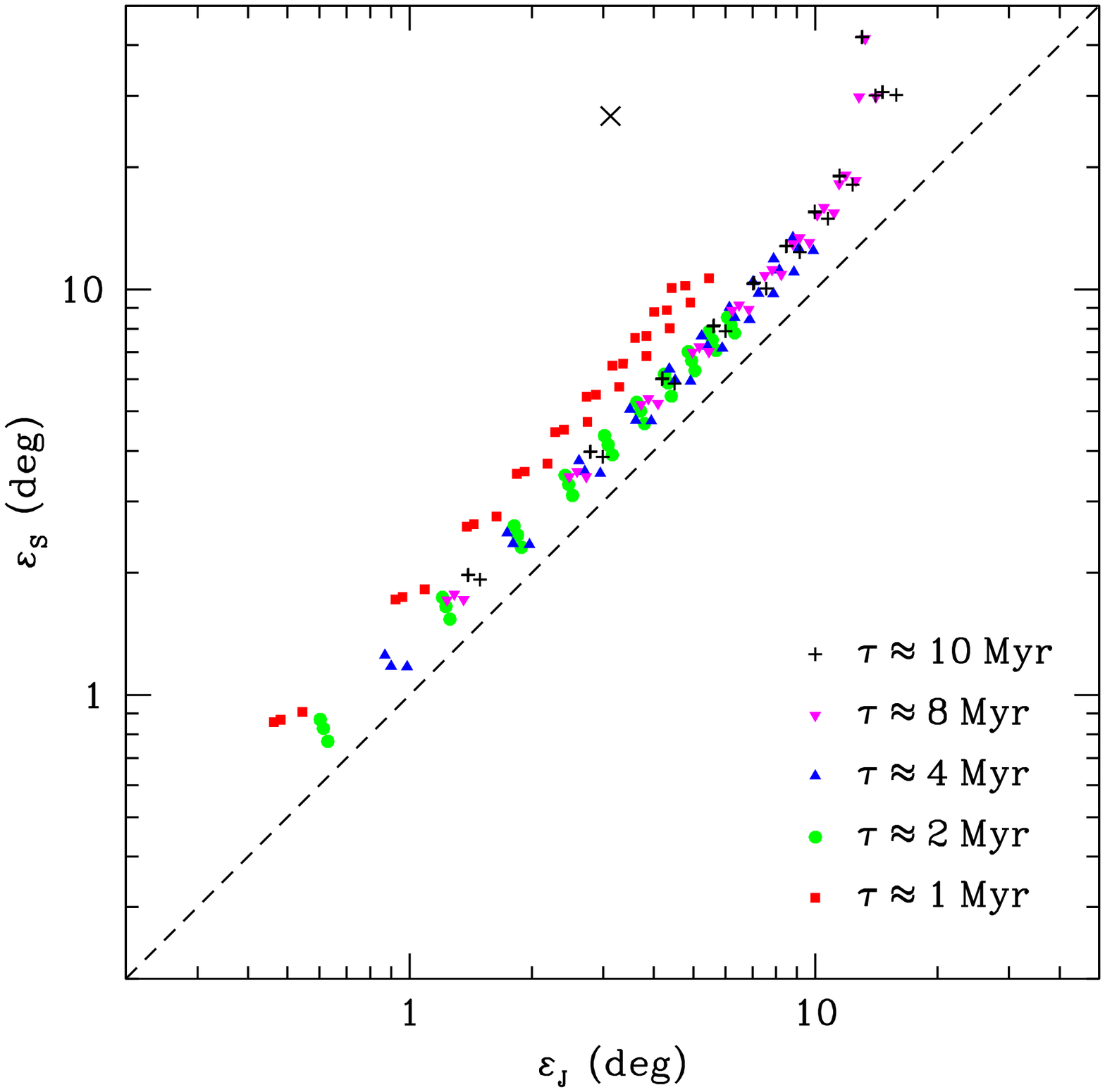}
\caption{ Same as Fig.\ref{fig:smooth_JS_overview}, but for secular simulations of the resonant Nice model with $i_{\mathrm{N}} =
1$--$10^\circ$.}
\label{fig:secular_nice_overview}
\end{figure}
Fig.~\ref{fig:secular_nice} shows the evolution of a resonant Nice model with $\tau= 10\Myr$ and initial $i_{\mathrm{N}} = 4^{\circ}$
{(the initial inclinations of the other planets are zero)}. This relatively high inclination of Neptune was chosen to mimic the typical
inclination value {during the encounter phase} from the $N$-body simulations. The Nice initial configuration is significantly more
compact, which has the effect of increasing the eigenfrequencies relative to the spin precession frequencies. In particular, the same
frequency that successfully tilted Saturn in the smooth migration scenario, $|s_8|$, is now larger than {\it Jupiter's} spin precession
frequency and must cross it to reach its final value near Saturn's. The situation for Jupiter is exactly comparable to the case of
tilting Saturn, and the only question is whether the product of the migration time scale and the inclination is large enough at the
time of the resonance crossing near $10\Myr$ to produce a significant Jovian obliquity. {(Recall that Jupiter can be tilted by a 
faster decrease in $|s_8|$ than Saturn; see Fig.~\ref{fig:saturn_only}.)} Unfortunately, the answer is yes. In this run, at the end we 
have $\epsJ \simeq 6^{\circ}$, which is much greater than the observed $3^{\circ}$, even though Jupiter does not stay in resonance 
until the end. Moreover, Saturn's obliquity of $\sim 8^{\circ}$ is not substantially larger. To increase $\epsS$ we require an 
increase in either $i_{\mathrm{N}}$ or $\tau$ or both, but since the same resonance is at work and the crossing times are only 
separated by $\sim\tau$, improving the Saturn value will make Jupiter's obliquity unrealistically large.
Fig.\ \ref{fig:secular_nice_overview} summarises the outcomes from a series of resonant Nice runs with $\tau$ varying from 1 to
$10\Myr$ and the initial $i_\mathrm{N}$ from 1--10$^{\circ}$. As feared, $\epsJ \sim \epsS$: the mean and standard deviation of
${\epsJ/\epsS = 0.67 \pm 0.12}$, whereas the observed ratio is ${0.11}$. The only runs which show $\epsS \geq {15^{\circ}}$ have
$10^{\circ} \leq \epsJ \leq 20^{\circ}$, and have long migration times (8 or $10\Myr$).

However, the secular models leave out many relevant dynamical effects. For example, we apply semi-major axis migration via our
prescriptions, and the real system will undergo a more sudden transition during resonance {crossing or} breaking and will be far more
stochastic. Perhaps Jupiter's spin-orbit resonance crossing happens much more quickly than Saturn's, preventing too large a tilt. In
the secular simulations we also let the inclinations undergo the usual evolution. Some experiments (not presented here) introducing
inclination damping proved unfruitful as the Jupiter crossing occurs first, and lowering the inclination enough to avoid tilting
Jupiter results in a far too small Saturnian obliquity.  In the real system, one could imagine that the inclinations are low during the
Jupiter spin-orbit resonance crossing and higher due to encounters when Saturn crosses, recovering the smooth migration model
behaviour. To determine whether these possibilities are realistic scenarios we must move beyond the secular models to more realistic
$N$-body models, and to those we now turn.

\subsection{\mbox{\boldmath $N$}-body Simulations}
\label{section:nbody}
We saw from the secular {simulations} that in the resonant Nice model the obliquities of Jupiter and Saturn are similar because of
$|s_8|$ crossing the spin precession frequencies of both planets. This did not occur in the smooth migration setup because of the wider
spacing of the planets.\footnote{{Although the classic Nice initial configurations \citep{tsi05} are also less compact than the
resonant ones, they have $|s_8|$ larger than Jupiter's spin precession frequency and have the same problem in the secular models.}}
We now present the results of our full $N$-body simulations. In what follows we need to quantify what we consider to be a successful
simulation that adequately reproduces the secular properties and spacing of the outer Solar System, {in addition to the obliquities of
Jupiter and Saturn}. To that end we identify several conditions that the system must adhere to. These are similar to the requirements 
of \cite{nes12} but with some modifications.
First, we need to have four giant planets (Jupiter, Saturn, and two ice giants) at the end of the simulation { -- criterion A of 
\cite{nes12}}. Second, the planets need to be within ${10\%}$ of their current orbits, i.e., $0.9 \le a_f/a_c \le 1.1$ where $a_f$ is 
the final semi-major axis of each planet at the end of the simulation, and $a_c$ is its current semi-major axis { -- similar to 
criterion B of \cite{nes12} which imposes a margin of 20\%.} Third, we impose a restriction on the difference in the 
longitudes of perihelion, $\Delta \varpi_{\rm JS} = \varpi_{\rm J} - \varpi_{\rm S}$. The current secular evolution of the 
eccentricities of Jupiter and Saturn is such that $\Delta\varpi_{\rm JS}$ circulates through $360^\circ$ and that the eccentricity of 
Jupiter (Saturn) reaches a maximum when $\Delta\varpi_{\rm JS} = 180^\circ$ ($0^\circ$). \cite{mor09} have shown that, at the very 
least, encounters between an ice giant {and} Saturn are needed to produce this secular behaviour. Thus, we require that $\Delta 
\varpi_{\rm JS}$ circulates. We used Fourier analysis over the last 5~Myr of the simulation to establish the behaviour of  $\Delta 
\varpi_{\rm JS}$. { This approach differs slightly from \cite{nes12} who keep track of the amplitude of the eccentricity eigenmode 
of Jupiter (their criterion C).} For the final obliquities, we require that $\epsJ < 5^\circ$ and $\epsS > 15^\circ$.

During the course of this investigation, we became aware of Jupiter often being too close to the Sun at the end of our $N$-body
simulations. For example, the mean and standard deviation of the final $a_{\rm J} = 4.98 \pm 0.21\au$ for the simulations of the
resonant Nice model with four planets at the end.
This affects not only whether the planets are within $10\%$ of their current orbits, but also potentially the obliquities of Jupiter
and Saturn, because the secular spin-orbit resonance problem is not scale free. The precession constants scale with the semi-major
axis of Jupiter as $\alpha \propto a_{\rm J}^{-3}$ while the secular eigenfrequencies scale as $s \propto
a_{\rm J}^{-3/2}$, if we fix the semi-major axis ratios. So the final $|s/\alpha|$ would be smaller by $\approx (3/2)
\Delta a_{\rm J}/a_{\rm J}$ if the final $a_{\rm J}$ is closer than the actual value $5.20\au$ by $\Delta a_{\rm J}$. Since the
initial $a_{\rm J}$ is in some sense arbitrary, we rescale each simulation so that the final $a_{\rm J} = 5.20\au$.
If the final $|s_8/\alpha_{\rm S}|< 1$ (i.e., there is resonance crossing) before rescaling and $|s_8/\alpha_{\rm S}|> 1$ (i.e., there 
is no resonance crossing) after rescaling, we use the obliquity of Saturn before the crossing in the original (unscaled) simulation as 
its final value. We apply the same procedure to $|s_7/\alpha_{\rm J}|$ for Jupiter.

\subsubsection{Final Orbits and Obliquities}
\begin{figure}[t]
\resizebox{\hsize}{!}{\includegraphics[angle=-90]{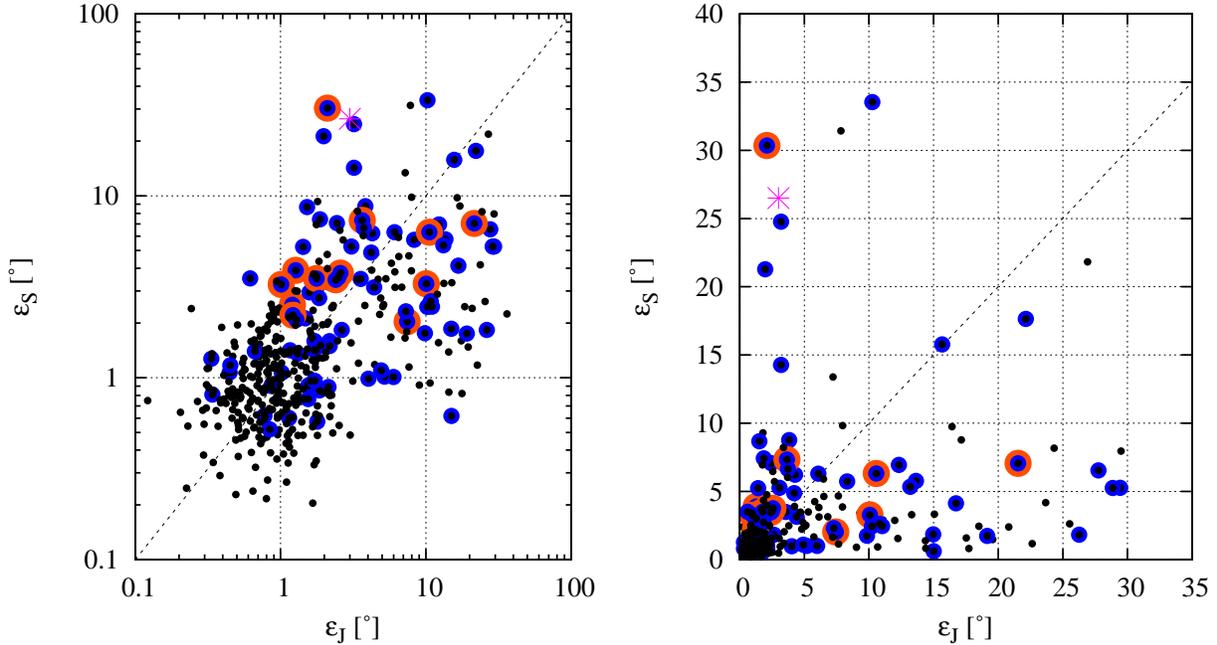}}
\caption{Final obliquities of Jupiter and Saturn in the $N$-body simulations of the resonant Nice model. Each panel depicts the same
outcome but with a different scale. Black dots denote simulations with four giant planets. Blue dots denote outcomes with four planets
in the right positions.
Finally, red dots denote outcomes that match all our {orbital} criteria. {The asterisk marks the observed configuration of Jupiter 
and Saturn.}}
\label{fig:epsjs4pl}
\end{figure}

First we analyse the outcome from the $N$-body simulations of the resonant Nice model, with the planets residing {initially} in the
3:2, 3:2, 4:3 resonant chain and a planetesimal disc of $50 M_\oplus$. In Fig.~\ref{fig:epsjs4pl} we plot the final $\epsJ$ versus
$\epsS$. The values of both $\epsJ$ and $\epsS$ are averages taken over the last 50~Myr in order to wash out any large-amplitude,
long-period oscillations, except, as noted earlier, when there is secular spin-orbit resonance crossing before rescaling but no
crossing after rescaling ($|s_7/\alpha_{\rm J}|$ for Jupiter and $|s_8/\alpha_{\rm S}|$ for Saturn), in which case the obliquity is
taken before the crossing. The left panel is a log-log version of the right panel. The small black dots depict the final obliquities
for simulations that have four planets {(but the planets may or may not be near their correct positions)}. The larger blue dots refer
to simulations where the planets are {also near} the correct positions 
(but $\Delta \varpi_{\rm JS}$ may librate or circulate). Finally, the red dots denote simulations that match all of our criteria {in 
orbital properties} for success. The dotted line denotes $\epsJ = \epsS$.

There are several visible trends in Figure \ref{fig:epsjs4pl}. First, the obliquities of both Saturn and Jupiter take on values ranging
from under 1$^\circ$ to over 30$^\circ$. They generally do not follow the line $\epsJ=\epsS$. Second, there are roughly equal numbers
of cases with $\epsJ > \epsS$ and  $\epsJ < \epsS$. We find that $\epsJ > \epsS$ 50\% of the time for all simulations that produce four
planets (irrespective of their final orbits), and $\epsJ > \epsS$ 51\% of the time for the cases when the planets are close to their
current orbits. However, the mean and standard deviation of $\epsJ/\epsS = 1.86 \pm 2.75$ and $2.24 \pm 3.40$ for each sample,
indicating that the distribution of $\epsJ/\epsS$ has a longer tail at large values. Compared to the secular simulations of the same
model in Section \ref{section:secular_nice}, the mean $\epsJ/\epsS$ is even further from the observed value of 0.11, but there is a
much larger spread in $\epsJ/\epsS$.

Unfortunately only very few simulations matched our criteria for being considered successful. A total of 84 runs {out of 1280} (6.6\%)
had the planets in the right positions. 
Of these only 13 have $\Delta \varpi_{\rm JS}$ circulate. Of these 13 only one has $\epsS>15^\circ$ and $\epsJ < 5^\circ$ {(see also 
Table \ref{tab:summary})}. However, it is not clear how essential of a constraint the circulation of $\Delta \varpi_{\rm JS}$ is. 
\cite{nes12} state that requiring the circulation of $\Delta \varpi_{\rm JS}$ is the real bottleneck in matching the current orbital 
architecture. Removing this restriction yields 2 additional good cases matching the planetary architecture and the obliquities. In 
summary, it appears that the 4-planet model has some difficulty reproducing the current Solar System. This fact was already 
established by \cite{nes12}, who reported similar statistics for the orbital constraints, but the probability is even lower ($\sim 
0.08$--$0.23\%$) when we consider the obliquities.


\begin{figure}[t]
\resizebox{\hsize}{!}{\includegraphics[angle=-90]{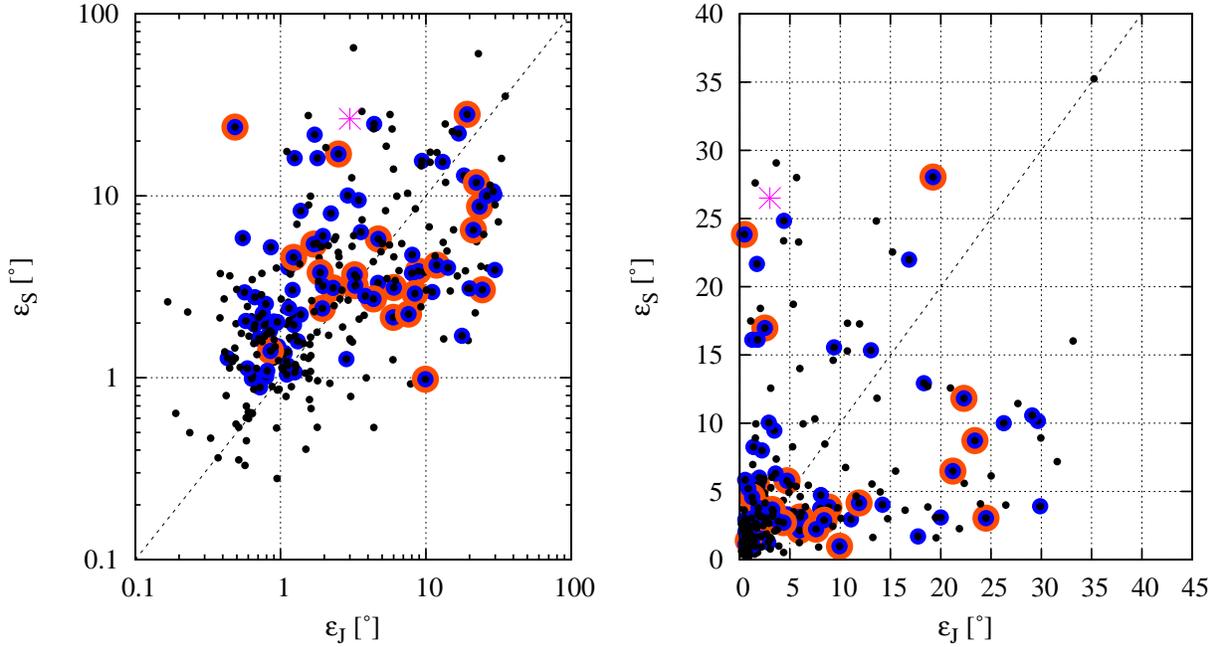}}
\caption{Same as Fig.~\ref{fig:epsjs4pl}, but for the $N$-body simulations of the compact (3:2 3:2 4:3 4:3) five planet model.}
\label{fig:epsjsMHL}
\end{figure}

Next we report on the outcome of the $N$-body simulations with five initial planets. Figure~\ref{fig:epsjsMHL} is similar to 
Fig.~\ref{fig:epsjs4pl} but now we focus on the compact five planet case with the initial configuration 3:2, 3:2, 4:3, 4:3 and a 
planetesimal disc of $35 M_\oplus$. The outcome is similar to, but slightly better than, the resonant Nice case with four planets: we 
have 6 cases where the giant planets are in the right positions and have their current obliquities, but in only 2 of these $\Delta 
\varpi_{\rm JS}$ circulates. There are fewer cases where Jupiter is tilted more than Saturn: $\epsJ > \epsS$ 37\% of the time for all 
simulations that produce four planets, and $\epsJ > \epsS$ 33\% of the time for the cases when the planets are close to their current 
orbits. But $\langle \epsJ/\epsS \rangle = 1.44 \pm 2.12$ for all final cases with four planets, and $1.45 \pm 2.02$ for when the 
final planets are close to their current orbits, again indicating that there is a longer tail at large values of $\epsJ/\epsS$.

\begin{figure}[t]
\resizebox{\hsize}{!}{\includegraphics[angle=-90]{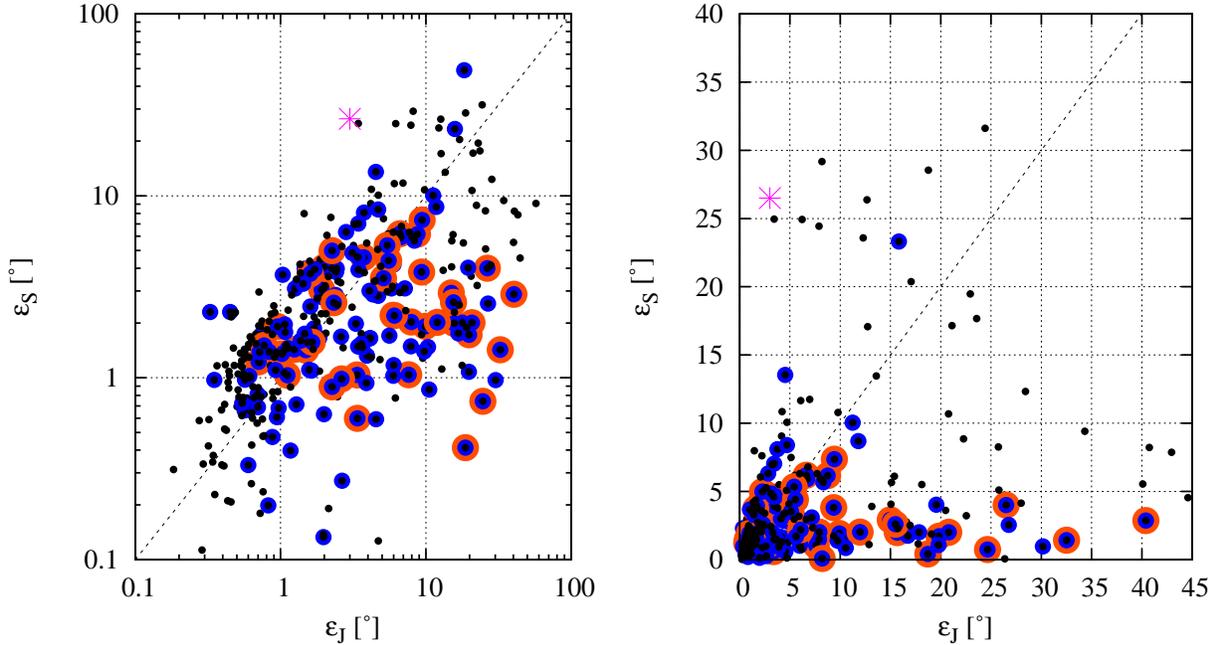}}
\caption{Same as Fig.~\ref{fig:epsjs4pl}, but for the $N$-body simulations of the loose (3:2 3:2 2:1 3:2) five planet model.}
\label{fig:epsjsDN}
\end{figure}

Finally, we turn to the loose five planet system, whose initial configuration is 3:2 3:2 2:1 3:2 and the outermost ice giant 
resides at $22.2\au$. The mass of the planetesimal disc is $20 M_\oplus$. We plot the final obliquities of Jupiter and Saturn in
Fig.~\ref{fig:epsjsDN}. The main difference with the same figures for the resonant Nice case (Fig.~\ref{fig:epsjs4pl}) and the compact
five planet configuration (Fig.~\ref{fig:epsjsMHL}) is an absence of high $\epsS$ at low $\epsJ$. Only one simulation with four
planets too far from their correct positions at the end has final $\epsS > 15^\circ$ and $\epsJ < 5^\circ$.  The fraction of
simulations with $\epsJ > \epsS$ is 45\% for all runs that have four planets at the end and 61\% for runs where the planets are near
their current orbits. We also find that $\langle \epsJ/\epsS \rangle = 3.98 \pm 3.45$ and $4.39 \pm 7.24$ for both samples,
respectively, so that this wider initial configuration of the planets leaves Jupiter's obliquity higher compared to Saturn's than the
compact one.

\subsubsection{Evolution examples}
\label{section:examples}

\begin{figure}
\resizebox{\hsize}{!}{\includegraphics[angle=-90]{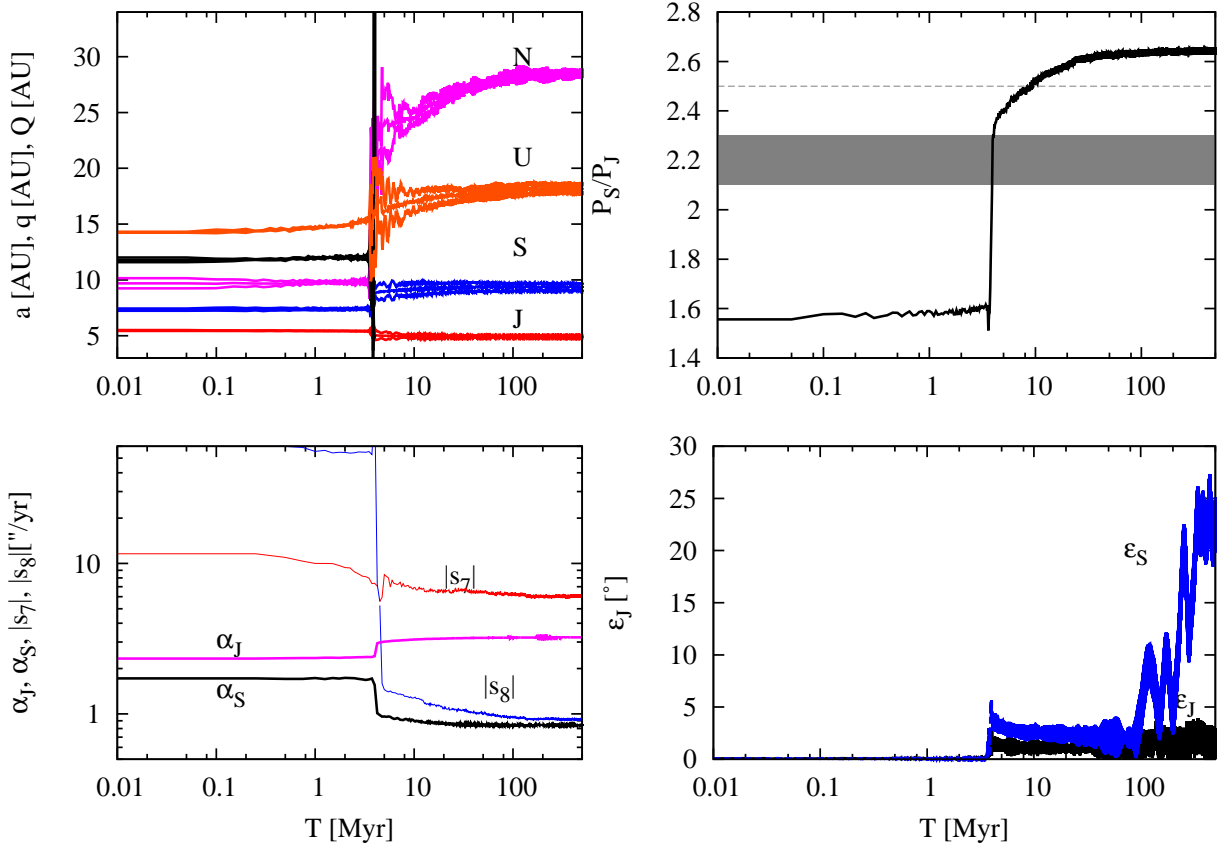}}
\caption{Evolution of a compact five planet case, where Saturn gets tilted and Jupiter does not. The top-left panel depicts the 
semi-major axis, perihelion and aphelion of the planets versus time. The letters and colour coding guide the eye. The top-right panel 
depicts the period ratio between Jupiter and Saturn versus time. The grey band is the region between 2.1 and 2.3 that needs to be 
crossed quickly. The dashed line shows the current value. The lower-left panel shows the spin precession constants of the gas giants 
and the vertical eigenfrequencies $s_7$ and $s_8$ as a function of time. Last, the bottom-right panel shows the obliquities of the gas 
giants with time.}
\label{fig:evohighs}
\end{figure}
\begin{figure}
\resizebox{\hsize}{!}{\includegraphics[angle=-90]{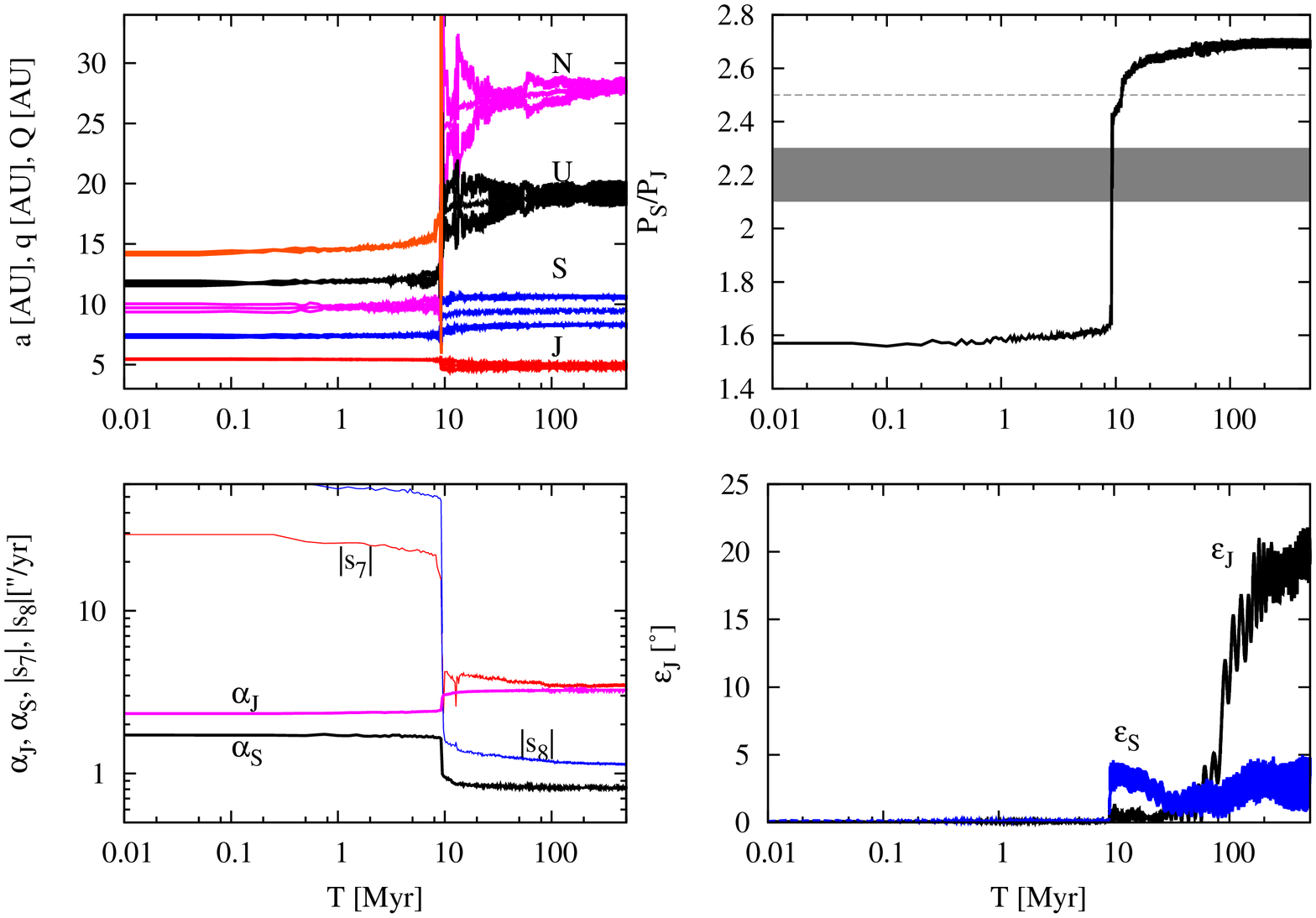}}
\caption{{Same as Fig.~\ref{fig:evohighs}, but for a case where Jupiter gets tilted but Saturn does not.}}
\label{fig:evohighj}
\end{figure}

We now illustrate the mechanisms behind the tilting of Jupiter and Saturn through some figures.
As was discussed earlier, tilting Saturn requires that $\alpha_{\rm S}$ resonates with $s_8$, or passes through this resonance. For
Jupiter, on the other hand, there are two ways to tilt it: passage through resonance with $s_8$ or $s_7$. Figure~\ref{fig:evohighs}
shows the evolution of the system that tilts both Jupiter and Saturn to their current obliquities in a simulation of the compact
five planet model. The top-left panel shows the perihelion distance, semi-major axis and aphelion distance of all planets as a 
function of time. To guide the eye each planet is identified by a different colour. The top-right panel depicts the period ratio 
between Jupiter and Saturn. The bottom-{right} panel depicts $\epsJ$ (black) and $\epsS$ (blue) and the bottom-{left} panel shows the 
evolution of $\alpha_{\rm J}$ (magenta), $\alpha_{\rm S}$ (black), $|s_7|$ (red) and $|s_8|$ (blue) as a function of time. When 
starting with three ice giants there is an extra vertical eigenfrequency that disappears when the third ice giant is ejected. The 
frequencies decrease with increasing semi-major axis and this property allowed us to keep track of which frequency corresponded to 
which planet and ultimately identify $s_7$ and $s_8$ belonging to the two surviving ice giants.

After approximately 3~Myr of evolution, the system becomes unstable. The innermost ice giant becomes Neptune, and the outermost ice 
giant evolves little and ends up being Uranus. The period ratio between the gas giants increases rapidly when Saturn scatters the lost 
ice giant and then evolves more slowly. The two remaining ice giants eventually settle into stable orbits at $t \approx 7\,$Myr and 
migrate towards their current orbits without undergoing encounters. The frequency $|s_8|$ crosses $\alpha_{\rm J}$ shortly after the 
system goes unstable but the crossing occurs during the jump and is too quick to allow capture in resonance. It then steadily 
approaches $\alpha_{\rm S}$ and after about 100~Myr it is close enough for Saturn to be caught in resonance. Saturn's obliquity 
increases and it stays in resonance until the end of the simulation. The resonant state is marked by the long-period oscillations in 
the obliquity of Saturn. At the same time $|s_7|$ {approaches} $\alpha_{\rm J}$ but {does not come} close enough for capture in 
resonance because Uranus stays fairly close to the Sun.

From Section~\ref{timescale} we know that a slow migration of Neptune is needed to tilt Saturn. From Fourier analysis we obtain for
the above simulation $I_{68} = 0.065^\circ$ and $\alpha_{\rm S} = 0.693\arcsec\,{\rm yr}^{-1}$, where $I_{68}$ is the
forced inclination in Saturn associated with $s_8$. Thus $\tau_{s,{\rm min}} \sim 10^9\,$yr, according to Figure
\ref{fig:saturn_only}{\it b} and equation (\ref{eq:tsmin}). In the simulation Neptune migrates outwards by 0.4~AU between $t=80$~Myr 
and $t=350$~Myr, and then essentially stops. This suggests that $\dot{a} \sim \Delta a/\Delta t \sim 2 \times 10^{-9}\,{\rm AU}\,{\rm 
yr}^{-1}$. The time scale for migration is then $t_{\rm mig} = a/\dot{a} \gg 10^9\,$yr and thus Saturn gets captured in resonance.


Unfortunately we also have many cases where Jupiter gets tilted significantly more than Saturn. In Fig.~\ref{fig:evohighj} we plot the
evolution of a simulation of the compact five planet model where Jupiter gets tilted but Saturn does not. The panels are the same as 
for Fig.~\ref{fig:evohighs}. The outermost ice giant is ejected when the system becomes unstable at $t \approx 9\,$Myr. From the
bottom-left panel we see that $\alpha_{\rm J}$ gets caught in resonance with $s_7$. Thus it appears that cases of high obliquity of
Jupiter are often caused by capture or passage through resonance with $s_7$ rather than $s_8$ because the latter occurs early in the
violent phase when the planets migrate fast. Applying Figure \ref{fig:saturn_only}{\it b} to Jupiter, we have $I_{57} = 0.072^\circ$ 
and $\alpha_{\rm J} = 3.24\arcsec\,{\rm yr}^{-1}$ and thus $\tau_{s,{\rm min}} \sim 200\,$Myr. Uranus migrates outward by 0.3~AU from 
$t=50$~Myr until $t=200$~Myr so that $t_{\rm mig} \gg 200\,$Myr. Thus Jupiter gets caught in the resonance.

\subsubsection{Spin-orbit resonance crossing}
\label{section:5plcases}

\begin{figure}[t]
\resizebox{\hsize}{!}{\includegraphics[angle=-90]{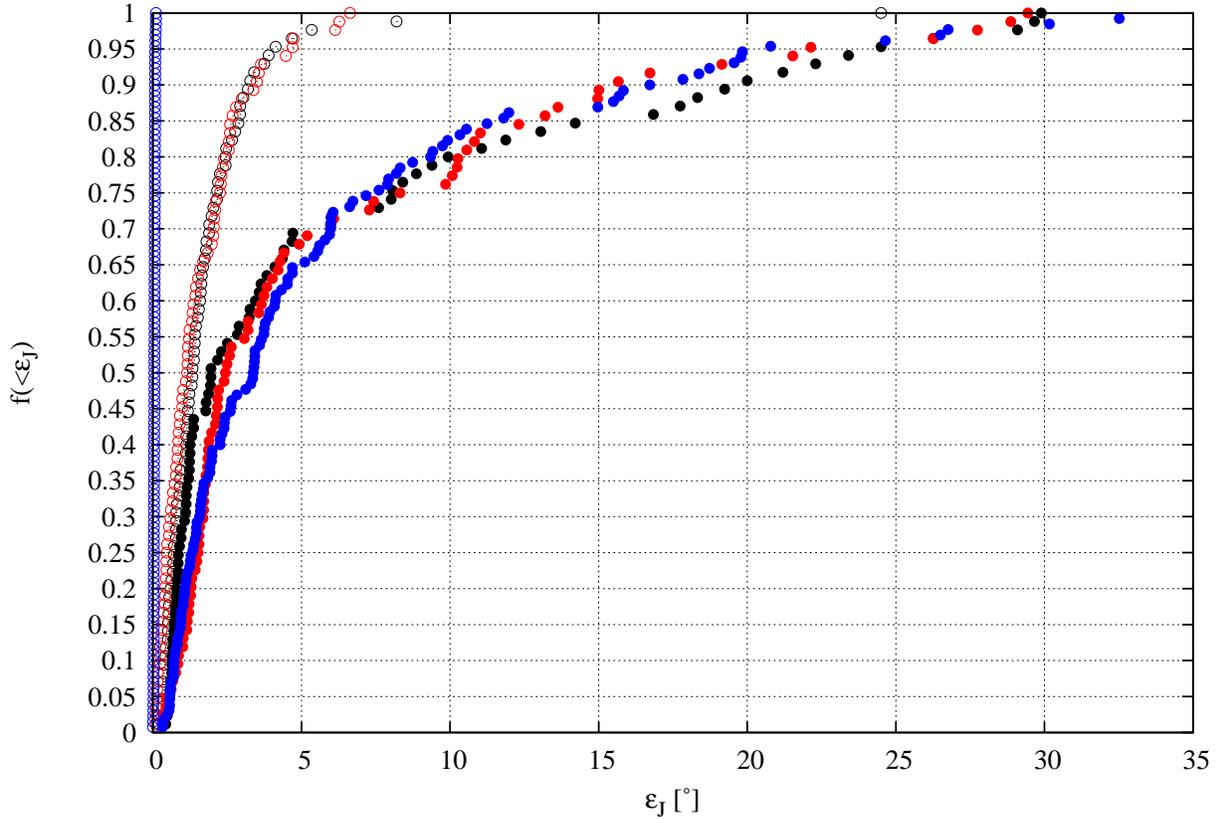}}
\caption{Cumulative distribution for Jupiter's obliquity to be less than $\epsJ$ for the $N$-body simulations with four planets in the
right positions at the end. The resonant Nice, compact five-planet, and loose five-planet models are shown in red, black, and blue,
respectively. The filled symbols show the cumulative distribution of the final obliquity, while the open symbols show the cumulative
distribution of the obliquity before any $|s_7|/\alpha_{\rm J}$ crossing (if one occurs).}
\label{fig:cumepsj}
\end{figure}

Figure \ref{fig:cumepsj} shows the cumulative distribution of Jupiter's final obliquity, compared with the cumulative distribution
of Jupiter's obliquity before any $|s_7|/\alpha_{\rm J}$ crossing (if one occurs), for the simulations with four planets in the right
positions at the end (blue dots in Figs.~\ref{fig:epsjs4pl}--\ref{fig:epsjsDN}). Except for one simulation in the compact five-planet 
model, all simulations have $\epsJ < 10^\circ$ before any $|s_7|/\alpha_{\rm J}$ crossing occurs, but at least $18\%$ of the 
simulations have final $\epsJ > 10^\circ$. This confirms that the cases of high Jupiter obliquity are often caused by capture or 
passage through resonance with $s_7$ instead of $s_8$. 

\begin{figure}[t]
\resizebox{\hsize}{!}{\includegraphics[angle=-90]{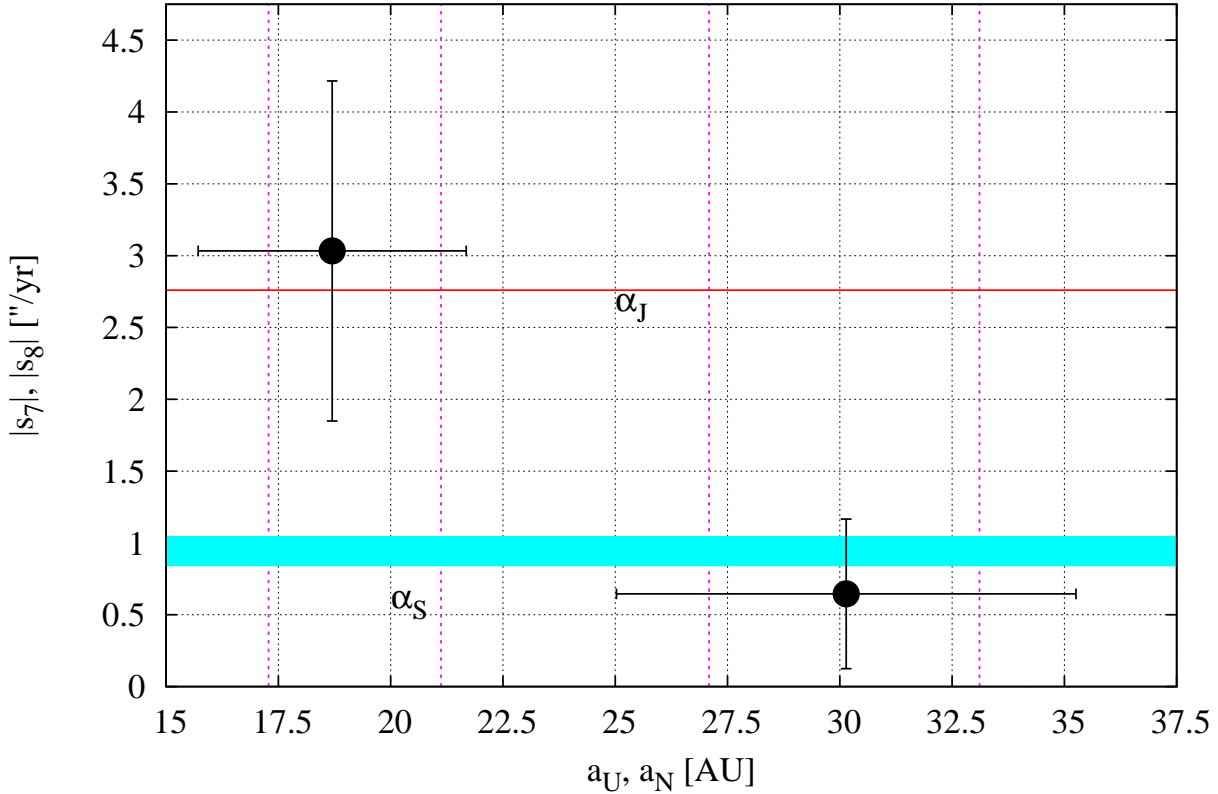}}
\caption{Final values of $|s_7|$ versus $a_{\rm U}$ and $|s_8|$ versus $a_{\rm N}$ for the $N$-body simulations with four planets in
the right positions at the end of the resonant Nice model (blue dots in Fig.~\ref{fig:epsjs4pl}), with the filled circles and error
bars showing the mean values and three standard deviations ($3\sigma$), respectively. The red line shows $\alpha_{\rm J}$, and the
cyan band shows the $\pm 3\sigma$ range of the final values of $\alpha_{\rm S}$. The vertical magenta dashed lines indicate the $\pm 
10\%$ range of the current $a_{\rm U}$ and $a_{\rm N}$.}
\label{fig:rat4pl}
\end{figure}

\begin{figure}[t]
\resizebox{\hsize}{!}{\includegraphics[angle=-90]{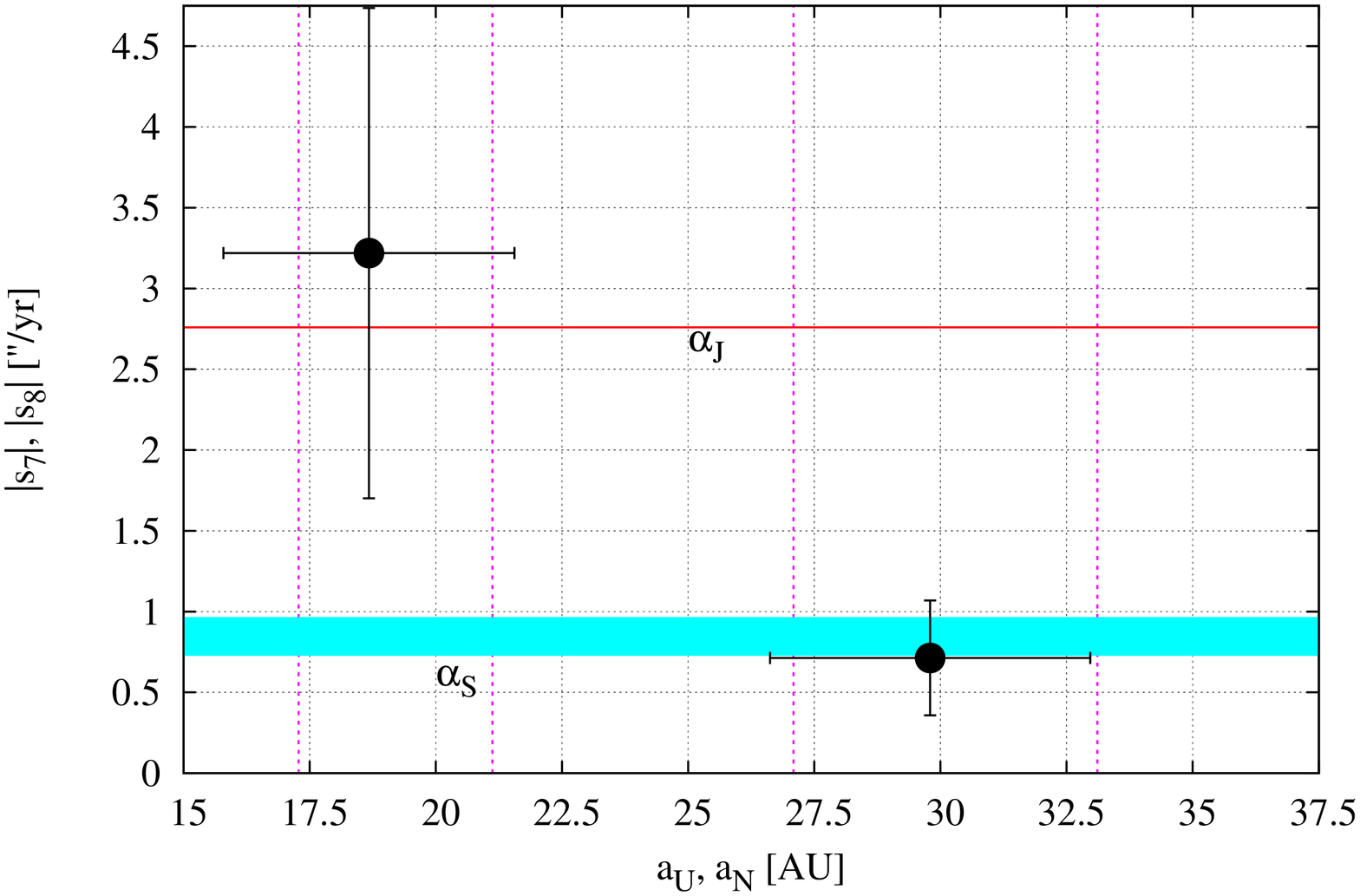}}
\caption{Same as Fig.~\ref{fig:rat4pl}, but for the $N$-body simulations with four planets in the right positions at the end of the
compact five planet model (blue dots in Fig.~\ref{fig:epsjsMHL}).}
\label{fig:ratMHL}
\end{figure}

\begin{figure}[t]
\resizebox{\hsize}{!}{\includegraphics[angle=-90]{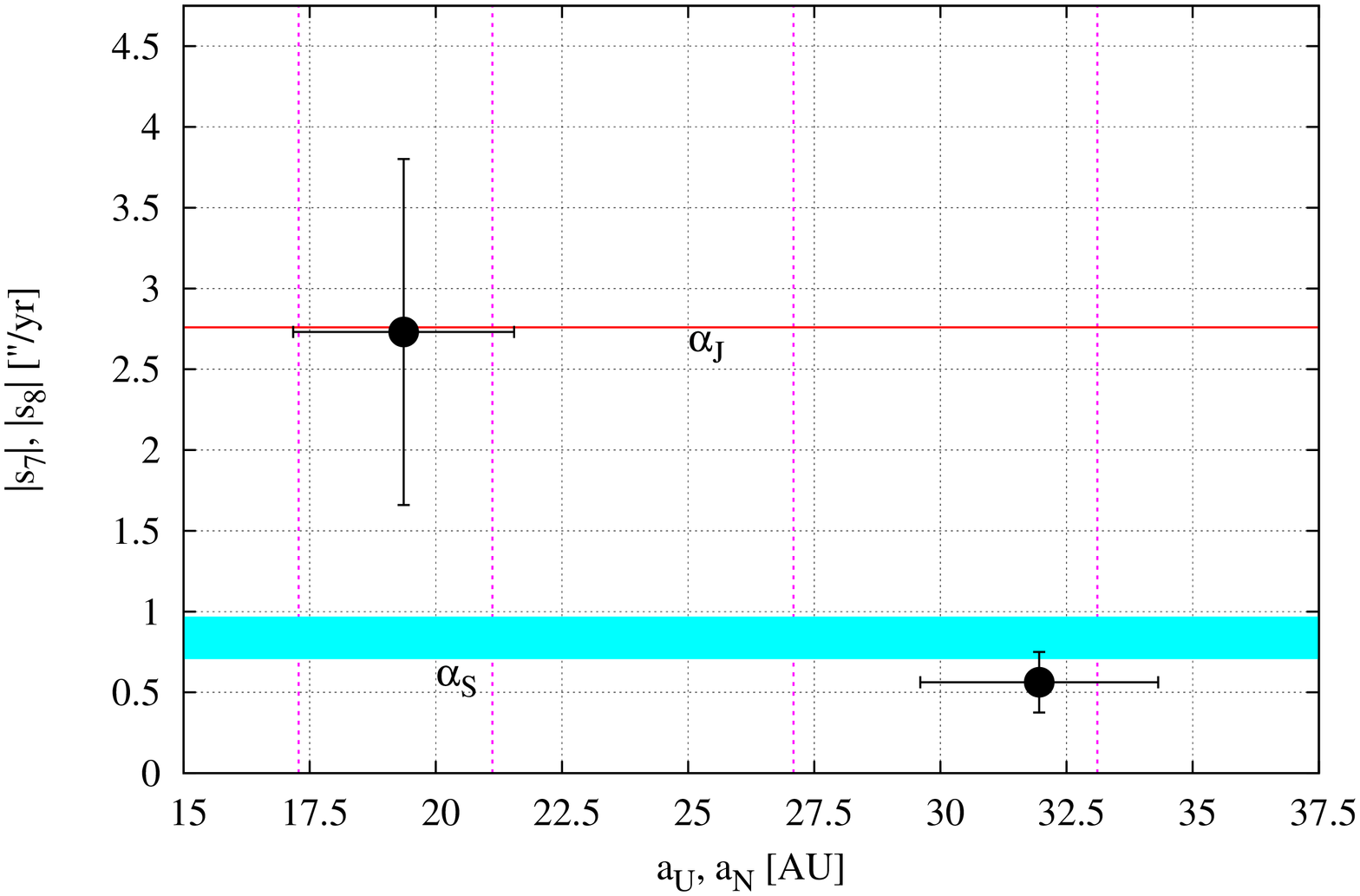}}
\caption{Same as Fig.~\ref{fig:rat4pl}, but for the $N$-body simulations with four planets in the right positions at the end of the
loose five planet model (blue dots in Fig.~\ref{fig:epsjsDN}).}
\label{fig:ratDN}
\end{figure}

To understand the differences in the distributions of the final obliquities of Jupiter and Saturn between the three models, we
compare in Figures \ref{fig:rat4pl}--\ref{fig:ratDN} the final values of $|s_7|$, $|s_8|$, $\alpha_{\rm J}$, and $\alpha_{\rm S}$
for the $N$-body simulations with four planets in the right positions at the end. In each figure, the filled circles and error bars 
show the mean values and three standard deviations ($3\sigma$) of the final values of $|s_7|$ versus $a_{\rm U}$ and $|s_8|$ versus 
$a_{\rm  N}$, with $s_7$ and $s_8$ from the Laplace-Lagrange secular theory. The red line shows the final $\alpha_{\rm J}$, which is 
fixed after rescaling the final $a_{\rm J}$ to $5.20\au$, and the cyan band shows the $\pm 3\sigma$ range of the final values of 
$\alpha_{\rm S}$. {The vertical magenta dashed lines indicate the $\pm 10\%$ range of the current $a_{\rm U}$ and $a_{\rm N}$.}

From Figure \ref{fig:rat4pl} for the resonant Nice model, we see that $|s_7|$ is about $0.7\sigma$ above $\alpha_{\rm J}$, which
means that $|s_7|$ crosses $\alpha_{\rm J}$ in about $24\%$ of the simulations with four planets in the right positions, if we
assume a normal distribution. This agrees with the $\sim 24\%$ of simulations with final $\epsJ > 10^\circ$ (see
Fig.~\ref{fig:cumepsj}), showing again that the cases where Jupiter has a high obliquity are most likely caused by a resonance 
crossing with $s_7$ instead of $s_8$. For the compact five planet model (Fig.~\ref{fig:ratMHL}), $|s_7|$ is further ($\sim 1\sigma$) 
above $\alpha_{\rm J}$, and there are fewer ($\sim 20\%$) simulations with $\epsJ > 10^\circ$. On the other hand, although more than 
half of the loose five planet simulations have final $|s_7| < \alpha_{\rm J}$ (see Fig.~\ref{fig:ratDN}), only $\sim 18\%$ of the 
simulations have $\epsJ > 10^\circ$. This suggests that, in this model, the $|s_7|/\alpha_{\rm J}$ resonance crossing often occurs 
when Uranus is migrating too quickly, which is consistent with the fact that Uranus is on average farther from the Sun than in the 
other two models.

Unfortunately, this problem of resonance crossing occurring too quickly also prevents the tilting of Saturn in many cases. For the 
resonant Nice model (Fig.~\ref{fig:rat4pl}), more than half of the simulations have $|s_8| < \alpha_{\rm S}$, but only $\sim 8\%$ of 
the simulations have $\epsS > 10^\circ$, indicating that the $|s_8|/\alpha_{\rm S}$ resonance crossing often occurs when Neptune is 
migrating too quickly. Although the compact five planet model has a similar mean value of $a_{\rm N}$, it has a narrower range in 
$a_{\rm N}$ and more overlap between $|s_8|$ and $\alpha_{\rm S}$ (see Fig.~\ref{fig:ratMHL}), which results in a higher fraction 
($\sim 19\%$) of simulations with $\epsS > 10^\circ$. Finally, in the loose five planet model (Fig.~\ref{fig:ratDN}), Neptune is on 
average too far from the Sun, and almost all of the simulations have $|s_8| < \alpha_{\rm S}$, but only $\sim 3\%$ of the simulations 
have $\epsS > 10^\circ$.

\subsubsection{Summary}
We summarize the outcome of our {$N$-body} simulations in Table~\ref{tab:summary}. The column $f_4$ is the probability of
the simulation having four giant planets after 500~Myr. The column $f_a$ is the probability that the system has four planets with each
planet within $10\%$ of its current orbit. The next column, $f_\varpi$, is the probability of having four planets close to their 
current orbits and $\Delta \varpi_{\rm JS}$ circulating. The last column, $f_\varepsilon$, is the probability that the system also has 
$\epsJ<5^\circ$ and $\epsS > 15^\circ$. For the resonant Nice and compact five planet models, we list two values for $f_\varepsilon$, 
with the lower value requiring the circulation of $\Delta\varpi_{\rm JS}$ and the higher value removing this restriction on the state 
of $\Delta\varpi_{\rm JS}$. For the loose five planet model, none of the simulations with four planets close to their current orbits
has $\epsJ<5^\circ$ and $\epsS > 15^\circ$, and there is only an upper limit on $f_\varepsilon$.

\cite{nes12} have examined the statistics of orbital properties for a large number of models, each with only $30$--$100$ $N$-body
simulations. In this paper, we have examined three representative models with 1280 simulations each. Our results on the orbital
statistics are much more accurate, and they are consistent with those found by \cite{nes12}. In particular, the loose five planet model
is the most successful in reproducing the positions of the giant planets and the circulation of $\Delta\varpi_{\rm JS}$. However, it
is clear from Table~\ref{tab:summary} that the loose five planet model fails to reproduce the obliquities of Jupiter and Saturn (with
the probability $f_\varepsilon < 0.08\%$) with the setup we have chosen. Thus the obliquities of Jupiter and Saturn provide a 
strong constraint that can alter the relative merits of different models and initial conditions.

In summary, it appears that both the resonant Nice and compact five planet models are able to reproduce simultaneously the current
orbital architecture of the giant planets and the obliquities of the gas giants, with the compact five planet model faring slightly
better. However, the probability is less than $0.5\%$, even if we lift the restriction on the state of $\Delta\varpi_{\rm JS}$. The
loose five planet model fails with a probability of less than $0.08\%$.

\begin{table}[t]
\caption{Statistics of $N$-body Simulation Results
\label{tab:summary}}
\begin{tabular}{cccccccc}\\
\tableline\tableline
 Model & $M_{\rm disc}$ & Initial $a_{\rm N}$ & $f_4$ & $f_a$ & $f_{\varpi}$ & $f_{\varepsilon}$\\
\tableline
 3:2 3:2 4:3 & 50 $M_{\oplus}$ & 11.6~AU & 35\% & 6.6\% & 1.0\% & 0.08--0.23\%\\
 3:2 3:2 4:3 4:3 & 35 $M_{\oplus}$ & 14.2~AU & 23\% & 6.7\% & 1.9\% & 0.16--0.47\%\\
 3:2 3:2 2:1 3:2 & 20 $M_{\oplus}$ & 22.2~AU & 27\% & 10\% & 3.3\% & $<$0.08\%\\
\tableline
\end{tabular}
\end{table}

\section{Discussion}
\label{section:discussion}
Even though the results presented in the previous section are extensive, they are by no means complete. In this section we discuss 
a few outstanding issues.

First, we did not impose any condition on the migration speed of the giant planets. \cite{nes12} imposed having the Saturn-Jupiter 
period ratio evolve from below 2.1 to beyond 2.3 in shorter than 1~Myr (their criterion D). We verified that all nine cases in our 
$N$-body simulations that reproduce the current obliquities of the gas giant have some sort of jump in the period ratio, although only 
one case from the compact five planet model had a jump in the right range, and is depicted in Fig.~\ref{fig:evohighs}. This would 
suggest that the compact five planet case is better overall at matching all constraints than the resonant Nice model.

Second, we only tested a single four planet and two five planet configurations because we opted to bracket the two end points of the 
initial spacing of the giant planets to determine whether the final outcomes favour one case over the other. The answer is clearly 
yes. We have had to run many more simulations per set of initial conditions than \cite{nes12} because we are trying to satisfy an 
additional constraint, namely reproducing the obliquities of the gas giants. Given that we found at most a few cases out of 1280 that 
match all constraints from one set of initial conditions, it becomes evident why we ran as many simulations as we did for each set 
of initial conditions, and why it is infeasible to test a larger sample of initial conditions.

Third, we generally did not change the mass of the planetesimal disc when running the simulations for one configuration. One could 
argue that if we increased the disc mass from 35~$M_{\oplus}$ to 50~$M_{\oplus}$ in the compact five planet case we may get a higher 
probability matching all constraints. However, increasing the disc mass from 35~$M_{\oplus}$ to 50~$M_{\oplus}$ would most likely damp 
the eccentricities of Jupiter and Saturn too much and make them inconsistent with their current values -- see \cite{nes12} for a more 
detailed discussion. We ran the same compact five planet model with a 50~$M_{\oplus}$ disc. Even though the probabilities $f_4$ to 
$f_{\varepsilon}$ show no improvement and are nearly identical to those of the compact five planet model with the 35~$M_{\oplus}$ 
disc, we found that the eccentricities of Jupiter and Saturn were inconsistent with their current values ($>2\sigma$ result). Thus, 
increasing the disc mass does not help our case and we do not report the results here. In addition, our current disc mass of 
35~$M_{\oplus}$ typically increases the period ratio of Jupiter and Saturn by 0.2 after the jump, which \cite{nes12} state may 
indicate 
the disc is too massive.

Another issue that we have not considered is the uncertainties in the spin precession constant $\alpha_{\rm J}$ and $\alpha_{\rm S}$.
\cite{vok15} have published an analysis based on one-planet simulations similar to those in Section \ref{timescale} (i.e.,
integration of spin evolution for a planet whose orbital nodal precession rate $|s|$ is decreased on timescale $\tau_s$). Their
analysis differs from ours in assuming that the final values of $s_7$ and $s_8$ are fixed at the observed Solar System ones, but they
considered a range of values for both $\alpha_{\rm J}$ and $\alpha_{\rm S}$. For the $|s_8|/\alpha_{\rm J}$ resonance crossing,
they find constraints on $I_{58}$ and $\tau_s$ for the resonance crossing to be too fast for capture and Jupiter's obliquity to be $\le
3^\circ$ after this crossing (which agrees with the open symbols in Fig.~\ref{fig:cumepsj}). Similarly, for the final 
approach of $s_7$ to its current value, they find constraints on $\alpha_{\rm S}$ and $\tau_s$ that yield Jupiter's final obliquity to 
be $\le 3^\circ$. For the final approach of $s_8$ to its current value, they considered different values of $\alpha_{\rm S}$ and 
$\tau_s$ and find that capture into the $|s_8|/\alpha_{\rm S}$ resonance requires large $\tau_s$ and low $\alpha_{\rm S}$ (including 
the value of $\alpha_{\rm S}$ adopted in the present paper). However, \cite{vok15} did not assess the probability that such conditions 
on the migration rates are satisfied in the various models of \cite{nes12}, nor did they assess the probability of these models giving 
the right values of $s_7$ and $s_8$, which we have done here for three specific configurations. Thus, while they suggest that the 
mechanism behind tilting Saturn is the resonance crossing between $s_8$ and $\alpha_S$, we show here how well this mechanism performs 
with full $N$-body simulations in several specific settings.

\begin{figure}[t]
\resizebox{\hsize}{!}{\includegraphics[angle=-90]{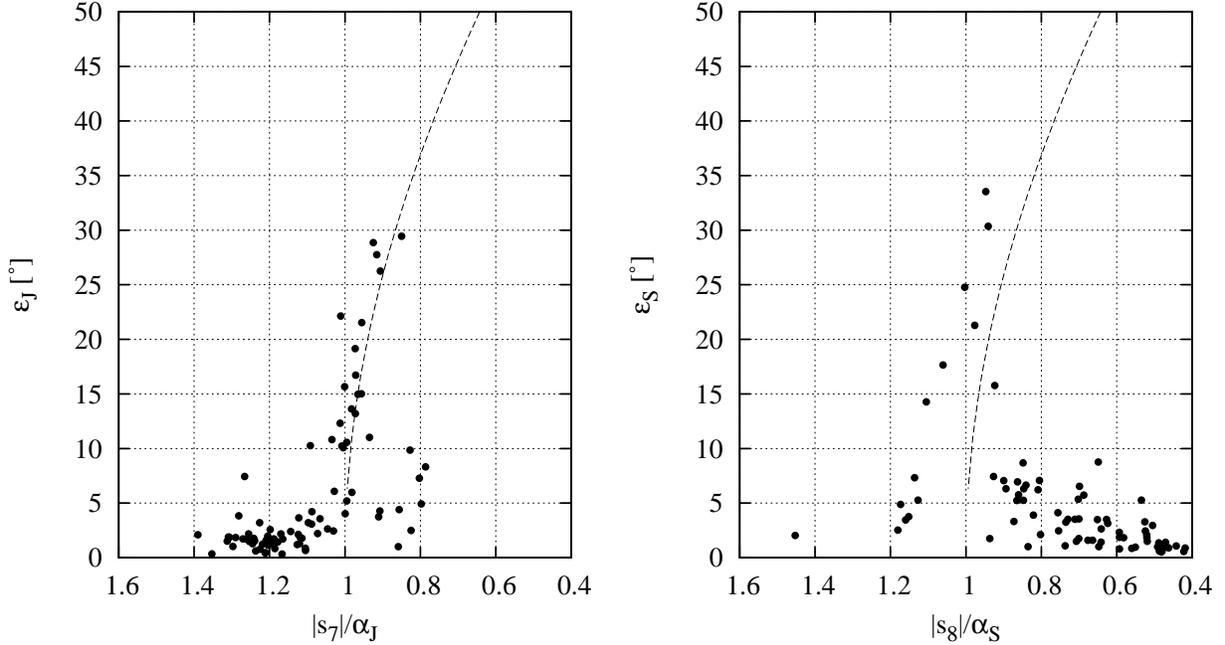}}
\caption{The final obliquity of Jupiter (left) and Saturn (right) vs $\vert s_7/\alpha_{\rm J}\vert$ (left) and $\vert s_8/\alpha_{\rm 
S}\vert$
(right) for the resonant Nice model. The dashed line shows $\varepsilon = \arccos(\vert s/\alpha\vert)$.}
\label{fig:rat4plmod}
\end{figure}

\begin{figure}[t]
\resizebox{\hsize}{!}{\includegraphics[angle=-90]{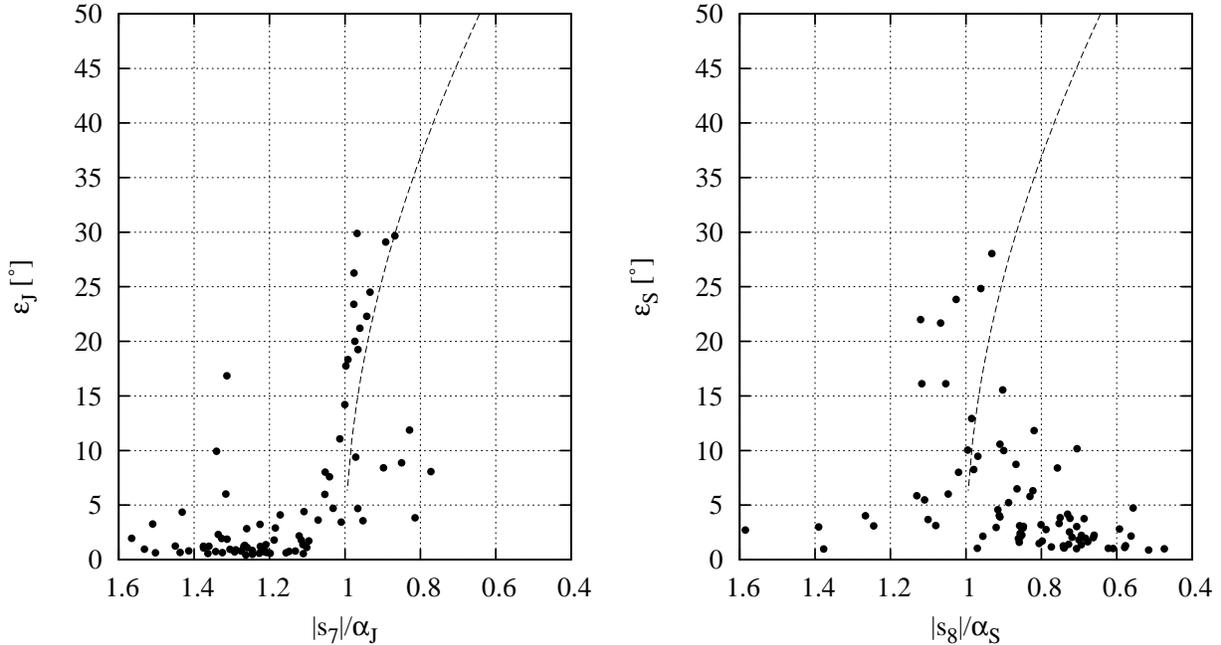}}
\caption{The final obliquity of Jupiter (left) and Saturn (right) vs $\vert s_7/\alpha_{\rm J}\vert$ (left) and $\vert s_8/\alpha_{\rm 
S}\vert$
(right) for the compact 5 planet model. The dashed line shows $\varepsilon = \arccos(\vert s/\alpha\vert)$.}
\label{fig:ratMHLmod}
\end{figure}

\begin{figure}[t]
\resizebox{\hsize}{!}{\includegraphics[angle=-90]{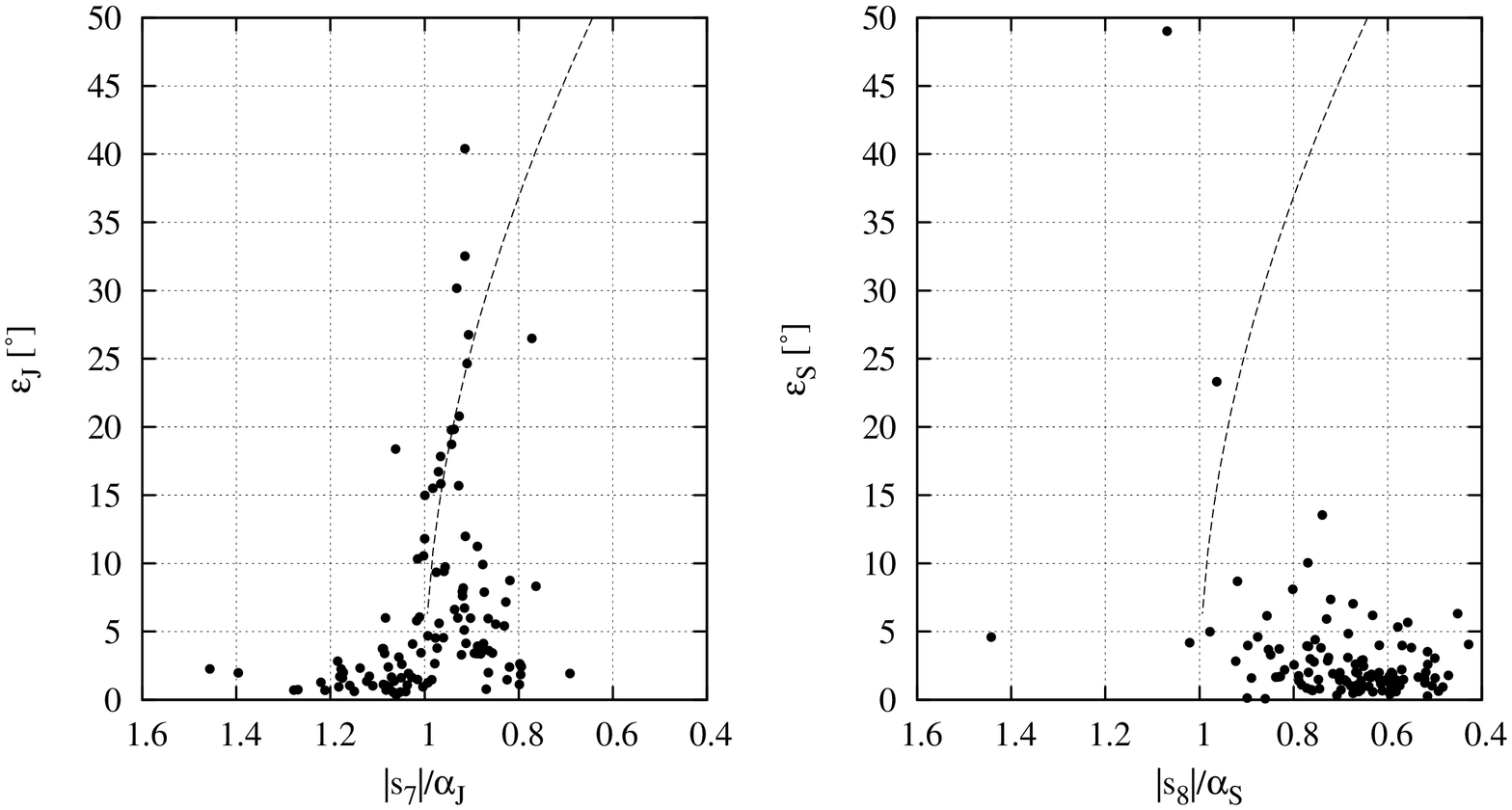}}
\caption{The final obliquity of Jupiter (left) and Saturn (right) vs $\vert s_7/\alpha_{\rm J}\vert$ (left) and $\vert s_8/\alpha_{\rm 
S}\vert$ 
(right) for the loose 5 planet model. The dashed line shows $\varepsilon = \arccos(\vert s/\alpha\vert)$.}
\label{fig:ratDNmod}
\end{figure}

A last point that requires elaboration is whether the obliquities of the gas giants can be used to constrain the migration of the 
giant planets at all. At issue is the following. The current solar system has $\epsJ \sim 3^\circ$ and $\epsS \sim 27^\circ$ and for 
both planets $\alpha_{\rm J} \cos \epsJ \sim \vert s_7 \vert$ and $\alpha_{\rm S} \cos \epsS \sim \vert s_8 \vert$. With a 10\% margin 
in the final semi-major axes of the planet one can compute the range of expected final obliquities of the gas giants assuming that 
during a resonance crossing there is perfect capture i.e. $\cos \epsJ' = \vert s_7'/\alpha_{\rm J}\vert$ if $\vert s_7'/\alpha_{\rm 
J}\vert \leq 1$, and $\cos \epsJ'=1$ otherwise. Here primed quantities are the values at the end of the simulation. Doing a series 
expansion of the Laplace-Lagrange theory around the current semi-major axes of the ice giants we find that $\epsJ' \in (0,44^\circ)$ 
and $\epsS' \in (0,53^\circ)$. With such a large range of final obliquities over a small range of final semi-major axes of the giant 
planets, one might argue that the final obliquities of the gas giants cannot be used to constrain the migration. 

However, this argument requires one assumption that does not stand up to scrutiny. Indeed, if there was perfect capture then in a plot 
of $\vert s/\alpha \vert$ vs $\varepsilon$ the data points should be clustered near $\varepsilon = \arccos(\vert s/\alpha\vert)$ 
when $\vert s/\alpha \vert \leq 1$ and near $\varepsilon =0$ otherwise. Figures~\ref{fig:rat4plmod}-\ref{fig:ratDNmod} depict the 
final obliquities of the gas giants as a function of $\vert s/\alpha\vert$. It is clear that this trend is only partly visible. We see 
some clustering of high $\epsJ$ near $\vert s_7/\alpha_{\rm J} \vert \sim 1$ -- and in some cases near $\vert s_7/\alpha_{\rm J} \vert 
> 1$ because 
of the inherent uncertainty in the Laplace-Lagrange model. For Jupiter the trend is more obvious than it is for Saturn, and the trend 
for Saturn is almost absent in the loose five planet model. This indicates that when a crossing does occur, capture often does 
not. In other words, the assumption that during a crossing there is perfect capture is clearly untrue.
 
So what are we to make of this? The answer is that for Saturn's obliquity to be where it is, two things need to be satisfied 
{\it simultaneously}: a) Neptune's migration needs to be slow enough so that $\tau \times i \gtrsim 30$~Myr~deg, and b) the spacing 
of the planets needs to be such that $\vert s_8/\alpha_{\rm S}\vert \approx 1$. We showed in Figs.~\ref{fig:rat4pl}-\ref{fig:ratDN} 
that the final semi-major axes of Uranus and Neptune (and indirectly, also the one of Saturn) are not the same for each set of 
simulations. Consequently, the final values of $s_7$ and $s_8$ are not the same either, despite the requirement of the final orbits 
being within 10\% of the current ones. As an example, for the resonant Nice model, the average and standard deviation of the final 
semi-major axes of Saturn, Uranus and Neptune in the cases where the planets are near their current orbits, are $\langle a_{\rm S} 
\rangle = 8.98 \pm 0.352$~AU, $\langle a_{\rm U} \rangle = 18.69 \pm 0.99$~AU and $\langle a_{\rm N} \rangle = 30.14 \pm 1.70$~AU. In 
contrast, for the compact five-planet model the values are $\langle a_{\rm S} \rangle = 9.33 \pm 0.451$~AU, $\langle a_{\rm U} \rangle 
= 18.69 \pm 0.956$~AU and $\langle a_{\rm N} \rangle = 29.74 \pm 1.08$~AU. Even though these are all within 10\% of the current values, 
the current spacing of the giant planets is better matched for the compact five planet case than for the resonant Nice case. It is the 
difference in the final spacing, coupled with the tail of Neptune's migration, that determines the obliquities of the gas giants and 
whether one model matches the observed constraints better than another. In light of this, we conclude that the obliquities of the gas 
giants can indeed be used to constrain the evolution and initial conditions of the giant planets.

\section{Summary and conclusions}
\label{section:disc_conclusions}
We have performed secular and $N$-body simulations tracking planetary obliquity to study constraints on migration histories of the
outer Solar System from both spin and {orbit}, covering the smooth migration scenario, the resonant Nice model and two five planet
scenarios. {The obliquities of Jupiter and Saturn provide a strong constraint on the models.} The {secular} results leave us in the
following situation: in the very simplest case, that of smooth migration, the \cite{wh04} secular spin-orbit resonance mechanism for
tilting Saturn's spin axis appears to work nicely if the product of the migration time scale and the orbital inclinations is
sufficiently large ($\tau \times i \ga 30$~Myr\,deg). On the other hand, the resonant Nice model, which is preferable on many
grounds (such as explaining the orbital properties of the giant planets, terrestrial planets, and main belt asteroids), is more
problematic, {because the secular eigenfrequency $s_8$ is initially above the spin precession frequency of Jupiter,
$\alpha_{\rm J}$}. The fundamental problem is that there are incompatible constraints. We need to migrate slowly ($\tau \ga 8\Myr$) at
typical inclinations to consistently tilt Saturn's spin axis by $> 15^{\circ}$, but for those same inclinations we need to migrate
quickly (say, $\tau \la 2\Myr$) to avoid tilting Jupiter by more than the observed $3^\circ$. We find that on average $\epsJ \sim
\epsS$ in our secular simulations of the resonant Nice model.

At the same time, the $N$-body simulations appear to tell a different story. The resonant Nice model is able to reproduce the
current orbital architecture of the giant planets and the obliquities of Jupiter and Saturn, but only with a small probability. 
The compact five planet model fares slightly better, but the loose five planet model has great difficulty reproducing the obliquities, 
even though it is the most successful in reproducing the orbital properties. However, it is possible that a slight change in the
initial conditions or the outer edge of the planetesimal disc could improve the outcome.

Ultimately the following needs to happen: (1) There needs to be fast migration during the encounter phase to avoid tilting Jupiter
through resonance passage with $s_8$. (2) Then a late, slow migration of Neptune to its current location could complete the task by
tilting Saturn {through the resonance with $s_8$.} (3) At the same time Uranus must stay close enough to the Sun to avoid tilting
Jupiter through the resonance with $s_7$. {Condition (1) is almost always satisfied in the $N$-body simulations. But the chances of
satisfying conditions (2) and (3) are limited, as the $|s_8|/\alpha_{\rm S}$ resonance crossing often occurs when Neptune is migrating
too fast, especially in the loose five-planet model, while Jupiter sometimes crosses the $|s_7|/\alpha_{\rm J}$ resonance. Thus the main
obstacle encountered to reproduce the obliquities of both Jupiter and Saturn is the final orbital spacing of the giant planets, coupled
with the tail of Neptune's migration. Both the
resonant Nice and compact five planet models appear to be able to overcome the obstacle, but with low probability.}\\
 
\footnotesize{
The authors are grateful for the support of Hong Kong RGC Grant HKU 7030/11P. We thank an anonymous reviewer for his/her valuable 
comments. We thank D.~Nesvorn\'y for providing one of his five planet configurations and fruitful discussions, A.~Morbidelli for 
comments on an earlier version, S.~J.~Peale for useful discussions during the early stages of this work, and D.~S.~McNeil for earlier 
work on this project. We are grateful for being able to use the TIARA grid computing cluster at the Institute for Astronomy and 
Astrophysics, Academia Sinica, Taiwan.}

\clearpage
\end{document}